\documentclass[aps,prl,showpacs,twocolumn,superscriptaddress]{revtex4}
\usepackage{graphicx}
\usepackage{bm}
\usepackage{amsmath,amssymb,amsfonts,bm}
\usepackage{epsfig}
\usepackage{epstopdf}
\usepackage{dcolumn}
\usepackage{grffile}
\usepackage{verbatim}
\usepackage{mathrsfs}
\usepackage[colorlinks=true,linkcolor=blue,citecolor=blue, urlcolor=blue]{hyperref}
\begin{document}


\title{Exact Solution to a Class of Generalized Kitaev Spin-$1/2$ Models in Arbitrary Dimensions}

\author{Jian-Jian Miao}
\affiliation{Kavli Institute for Theoretical Sciences, University of Chinese Academy of Sciences, Beijing 100190, China}

\author{Hui-Ke Jin}
\affiliation{Department of Physics, Zhejiang University, Hangzhou 310027, China}

\author{Fu-Chun Zhang}
\affiliation{Kavli Institute for Theoretical Sciences, University of Chinese Academy of Sciences, Beijing 100190, China}
\affiliation{CAS Center for Excellence in Topological Quantum Computation, University of Chinese Academy of Sciences, Beijing 100190, China}
\affiliation{Collaborative Innovation Center of Advanced Microstructures, Nanjing University, Nanjing 210093, China}

\author{Yi Zhou}
\affiliation{Department of Physics, Zhejiang University, Hangzhou 310027, China}
\affiliation{CAS Center for Excellence in Topological Quantum Computation, University of Chinese Academy of Sciences, Beijing 100190, China}
\affiliation{Collaborative Innovation Center of Advanced Microstructures, Nanjing University, Nanjing 210093, China}

\date{\today}

\begin{abstract}

We construct a class of exactly solvable generalized Kitaev spin-$1/2$ models in arbitrary dimensions, which is beyond the category of quantum compass models. 
The Jordan-Wigner transformation is employed to prove the exact solvability. 
An exactly solvable quantum spin-$1/2$ models can be mapped to a gas of free Majorana fermions coupled to static $Z_2$ gauge fields. 
We classify these exactly solvable models according to their parent models. Any model belonging to this class can be generated by one of the parent models.
For illustration, a two dimensional ($2D$) tetragon-octagon model and a three dimensional ($3D$) $xy$ bond model are studied.

\end{abstract}

\maketitle


Kitaev honeycomb model\cite{Kitaev} has attracted a lot of attention for it is simple in form but rich in physics.
A variety of research fields in physics, including topological phases of matter, strongly correlated electrons and topological quantum computation, converge in this model. 
Unexpectedly, such an interacting two-dimensional ($2D$) quantum spin model can be solved exactly. Hence we can explore the physics of the model without the interruption from various approximation methods, 
which are inevitably adopted to handle nonintegrable quantum many-body systems. Moreover the model gives rise to a topologically nontrivial phase hosting non-Abelian anyons, 
which can be manipulated for fault tolerant quantum computation\cite{Kitaev_03}. 
The most serious obstacle to build a quantum computer comes from the decoherence. Topological quantum computation overcomes this problem by utilizing the exotic topological properties of topological phases of matter\cite{TQCRMP}. 
The unitary evolutions of the qubits are performed by braiding the non-Abelian anyons, which is immune to any local perturbations.
A famous and classic example of topological phases of matter is the fractional quantum Hall effect\cite{Laughlin}. 
Especially the $\nu=5/2$ fractional quantum Hall state, which is a promising non-Abelian state with fractional excitations and non-Abelian anyons\cite{Moore}, 
has been proposed as the archetype for fault tolerant quantum computation. 
Besides the $\nu=5/2$ fractional quantum Hall state, there exist other candidate platforms for topological quantum computation as well\cite{TQCRMP}.

An exact solution to a quantum many-body system in dimensions greater than one $(D > 1)$ is rare and always sheds light on understanding the nature of strongly correlated systems. 
Kitaev honeycomb model is solved exactly by mapping the spin-$1/2$ model to a free Majorana fermions coupled to a static $Z_2$ gauge field. 
The exact solvability originates from the infinite number of conserved quantities in the thermodynamic limit.
In addition to the exact solution through four Majorana decomposition, which is pioneered by Kitaev himself, 
it was also found that the honeycomb spin-$1/2$ model can be exactly solved with the help of Jordan-Wigner transformation\cite{Feng,Chen,Chen-Nussinov}. 
This elegant method enables a fermionization of the spin model without redundant degrees of freedom, and allows it to be mapped to a $p$-wave-type Hubbard-BCS pairing problem\cite{Ng}. 
Besides the four Majorana decomposition and the Jordan-Wigner transformation, Nussinov and Ortiz also proposed another aspect of the exact solvability by focusing on the bond algebra\cite{Nussinov_09}. 
Because of the significance of Kitaev honeycomb model in physics, a lot of efforts are devoted to search for its generalizations with exact solvability, 
which include other $(2D)$ models\cite{yao2007,yang,Kells,baskaran2009,tikhonov2010}, three dimensional $(3D)$ models\cite{si,ryu,mandal,Hermanns_14,Kimchi,Trebst15,Hermanns,Nasu}, 
the models with multiple-spin interactions\cite{lee2007,yu2008}, $SU(2)$ invariant models\cite{wang,yao2011,lai2011} and higher spin models\cite{yao2009,wu,chern,chua,nakai,nussinov2013}.

In this paper, we construct a class of generalized Kitaev spin-$1/2$ models in arbitrary dimensions, which can be solved exactly with the aid of the Jordan-Wigner transformation. 
The model construction starts with a $d$-dimensional cube. We allocate various bonds on some links and erase the other links to obtain a new lattice and an exactly solvable model.
The allocation of bonds is subject to two elementary rules and several supplementary rules. We find that there exist a dual spin model to each constructed spin model.

\section{Model Hamiltonian}
Consider a $d$-dimensional cube, for $d=2,3,4$,..., it is square, cubic, hypercubic,..., lattice respectively.
Labeling each site as $n=(n_1,n_2,\cdots,n_d)$, where $1\leq n_{j}\leq L_{j}$ and $L_{j}$ is the length along $j$th-direction, $j=1,\cdots,d$. 
Assigning each site $n$ a number $\mathscr{N}=n_{1}+\sum_{j=2}^{d}\left(n_{j}-1\right)\left(\prod_{l=1}^{j-1}L_{l}\right)$, one is able to sort all the sites as follows: if $\mathscr{N}<\mathscr{M}$, then $n<m$. 
A {\em local link} is defined as a pair of sites $(n,m)$ with $\sum_{j=1}^{d}\left|n_{j}-m_{j}\right|=1$, while a {\em nonlocal link} $(n,m)$ is given by $\sum_{j=1}^{d}\left|n_{j}-m_{j}\right|>1$. 
The Hamiltonian of a generalized Kitaev spin-$1/2$ model consists of three parts,
\begin{equation}\label{eq:full_Ham}
H=H_{local}^{\left(2\right)}+H_{nonlocal}^{\left(2\right)}+H_{nonlocal}^{\left(M\right)},
\end{equation}
where $H_{local}^{\left(2\right)}$, $H_{nonlocal}^{\left(2\right)}$ and $H_{nonlocal}^{\left(M\right)}$ represent local two-spin interaction, 
nonlocal two spin-interaction and nonlocal multiple-spin interaction respectively.
$H_{local}^{\left(2\right)}$ describes two-spin interactions defined on the local links,
\begin{equation}
J_{nn+\hat{1}}^{\alpha\beta}\sigma_{n}^{\alpha}\sigma_{n+\hat{1}}^{\beta}
\end{equation}
and
\begin{equation}
J_{nm}^{zz}\sigma_{n}^{z}\sigma_{m}^{z},\label{eq:Jzz1}
\end{equation}
where $\alpha,\beta=x,y$ and $\sigma_{n}^{\alpha}$ are Pauli matrices at site $n$ and $m=n+\hat{j}$ 
with $\hat{j}$ the unit vector along the $j$-th direction and $j=1,\cdots,d$. 
$J_{nm}^{\alpha\beta}$ and $J_{nm}^{zz}$ are coupling constants. 
Similarly, $H_{nonlocal}^{\left(2\right)}$ describes two-spin interactions defined on the nonlocal links,
\begin{equation}
J_{nm}^{zz}\sigma_{n}^{z}\sigma_{m}^{z},
\end{equation}
and $H_{nonlocal}^{\left(M\right)}$ describes multiple-spin interactions defined on the nonlocal links,
\begin{equation}\label{eq:Ham_nonlocal}
J_{nm}^{\alpha\beta}\sigma_{n}^{\alpha}\left[\prod_{n<l<m}(-\sigma_{l}^{z})\right]\sigma_{m}^{\beta},
\end{equation}
through the string operator $\prod_{n<l<m}(-\sigma_{l}^{z})$ linking the sites $n$ and $m$, where $\alpha,\beta=x,y$ as well, 
and the extra minus sign is introduced in $(-\sigma_{l}^{z})$ for later convenience. In the above definition, we always keep $n<m$. 
So far we have five types of interactions, which can be distinguished by corresponding coupling constants $J_{nm}^{xx}$, $J_{nm}^{yy}$, $J_{nm}^{zz}$, $J_{nm}^{xy}$ and $J_{nm}^{yx}$. 
Hereafter we shall call them $x$-bond, $y$-bond, $z$-bond, $xy$-bond and $yx$-bond respectively.
Suppose we have a specific pattern of dividing all the sites to be white (w) or black (b). Indeed, such a pattern can be chosen by dividing the lattice into two arbitrary sublattices. 
The exact solvable model described by Eq.~\eqref{eq:full_Ham}-Eq.~\eqref{eq:Ham_nonlocal} can be constructed by allocating various bonds on the lattice, subject to two {\em elementary rules} as follows:
\begin{enumerate}
\item A (local or nonlocal) $x$-bond is allocated on a link $(n,m)$ with $n\in w$ and $m\in b$; a $y$-bond is allocated for $n\in b$ and $m\in w$; 
an $xy$-bond is allocated for $n\in w$ and $m\in w$; and a $yx$-bond is allocated for $n\in b$ and $m\in b$.
\item Different $z$-bonds are not allowed to share the same site.
\end{enumerate}
Here $w$ and $b$ refer to white and black sublattice respectively. 
The construction is to allocate various bonds on the $d$-dimensional cube lattice to form a connected graph. 
Different allocations give rise to different models.
Note that this construction allows some orphan sites which do not connect to any other sites through local or nonlocal bonds.
Then we just omit these isolated sites and obtain a new lattice from the original $d$-dimensional cube.

\section{Exact solvability}
We shall prove the exact solvability of the generalized Kitaev spin-$1/2$ models with the help of the Jordan-Wigner transformation\cite{Jordan-Wigner},
\begin{subequations}
\begin{eqnarray}
\sigma_{m}^{+} & = & c_{m}^{\dagger}e^{i\pi\left(\sum_{l<m}\hat{n}_{l}\right)},\\
\sigma_{m}^{z} & =& 2\hat{n}_{m}-1,
\end{eqnarray}
\end{subequations}
where $\sigma_{m}^{+}=\frac{1}{2}(\sigma_{m}^{x}+i\sigma_{m}^{y})$ is the spin raising operator, $c_{m}^{\dagger}$ is the creation operator for the spinless fermion at site $m$, 
and $\hat{n}_{m}=c_{m}^{\dagger}c_{m}$ is the fermion occupation number operator at site $m$. 
Then we decompose each complex fermion $c_n$ into two Majorana fermions $\eta_{n}$ and $\gamma_{n}$ as follows: (1) for $n\in w$, $\eta_{n}=c_{n}^{\dagger}+c_{n}$ 
and $\gamma_{n}=i\left(c_{n}^{\dagger}-c_{n}\right)$, (2) for $n\in b$, $\eta_{n}=i\left(c_{n}^{\dagger}-c_{n}\right)$ and $\gamma_{n}=c_{n}^{\dagger}+c_{n}$.
After the Jordan-Wigner transformation, allowed local $x$-bonds, $y$-bonds, $xy$-bonds and $yx$-bonds become
\begin{subequations}\label{eq:gamma}
\begin{eqnarray}
\sigma_{n\in w}^{x}\sigma_{n+\hat{1}\in b}^{x} & = &-i\gamma_{n}\gamma_{n+\hat{1}},\\
\sigma_{n\in b}^{y}\sigma_{n+\hat{1}\in w}^{y} & = & i\gamma_{n}\gamma_{n+\hat{1}},\\
\sigma_{n\in w}^{x}\sigma_{n+\hat{1}\in w}^{y} & = & -i\gamma_{n}\gamma_{n+\hat{1}},\\
\sigma_{n\in b}^{y}\sigma_{n+\hat{1}\in b}^{x} & = & i\gamma_{n}\gamma_{n+\hat{1}},
\end{eqnarray}
nonlocal $x$-bonds, $y$-bonds, $xy$-bonds and $yx$-bonds become
\begin{eqnarray}
\sigma_{n\in w}^{x}\left[\prod_{n<l<m}(-\sigma_{l}^{z})\right]\sigma_{m\in b}^{x} & = &-i\gamma_{n}\gamma_{m},\\
\sigma_{n\in b}^{y}\left[\prod_{n<l<m}(-\sigma_{l}^{z})\right]\sigma_{m\in w}^{y} & = &i\gamma_{n}\gamma_{m},\\
\sigma_{n\in w}^{x}\left[\prod_{n<l<m}(-\sigma_{l}^{z})\right]\sigma_{m\in w}^{y} & = &-i\gamma_{n}\gamma_{m},\\
\sigma_{n\in b}^{y}\left[\prod_{n<l<m}(-\sigma_{l}^{z})\right]\sigma_{m\in b}^{x} & = &i\gamma_{n}\gamma_{m},
\end{eqnarray}
and allowed $z$-bonds become
\begin{equation}
\sigma_{n}^{z}\sigma_{m}^{z}=i\hat{D}_{nm}\gamma_{n}\gamma_{nm},
\end{equation}
\end{subequations}
where $\hat{D}_{nm}=\pm i\eta_{n}\eta_{m}$ is defined along a $z$-bond only. 
The sign is $-$ when $n$ and $m$ belong to the same sublattice, while it is $+$ when $n$ and $m$ belong to the opposite sublattice.
Because of {\em rule} $2$, $\hat{D}_{nm}$ commute with each other and with the Hamiltonian $H$. 
Hence $\hat{D}_{nm}$ is a constant of motion and can be viewed as a static local $Z_2$ gauge field since $\hat{D}_{nm}^{2}=1$. 
To go further, we can replace the operator $\hat{D}_{nm}$ by its eigenvalues $D_{nm}=\pm1$. 
The eigenstates of the Hamiltonian can be divided into different sectors of total Hilbert space according to the sets of eigenvalues $\left\{ D_{nm}\right\}$. 
In each sector, all the allowed spin interactions are transformed to quadratic Majorana fermion terms and the Hamiltonian is exactly diagonalizable.


\section{Lift possible local degeneracy}
It is indicated in the proof of exact solvability that the fermionized Hamiltonian has the following structure,
\begin{equation}
H=H_{\gamma}\otimes H_{\eta},
\end{equation}
where $H_{\gamma}$ consists of quadratic $\gamma$ Majorana fermion terms only and $H_{\eta}$ consists of quadratic $\eta$ Majorana fermion terms only. 
It may occur that some $\eta_{n}$ do not show up explicitly in the fermionized Hamiltonian $H_{\eta}$ at all. 
This will happen if the site $n$ does not connect to any other sites through $z$-bond (but may connect through other types of bonds). 
These localized $\eta_{n}$ will give rise to {\em local degeneracy} in these constructed spin-$1/2$ models.

In order to lift the local degeneracy, we need to couple these isolated $\eta_{n}$ with each other to form a connected graph through {\em additional} two-spin and/or multiple-spin interactions, which is beyond the two elementary rules. 
This can be done without spoiling the exact solvability noting that the following two-spin and multiple-spin interactions can be fermionized by Jordan-Wigner transformation to quadratic $\eta$ Majorana fermion terms,
\begin{subequations}\label{eq:eta}
\begin{eqnarray}
\sigma_{n\in b}^{x}\left[\prod_{n<l<m}(-\sigma_{l}^{z})\right]\sigma_{m\in w}^{x} & = & -i\eta_{n}\eta_{m},\\
\sigma_{n\in w}^{y}\left[\prod_{n<l<m}(-\sigma_{l}^{z})\right]\sigma_{m\in b}^{y} & = & i\eta_{n}\eta_{m},\\
\sigma_{n\in b}^{x}\left[\prod_{n<l<m}(-\sigma_{l}^{z})\right]\sigma_{m\in b}^{y} & = & -i\eta_{n}\eta_{m},\\
\sigma_{n\in w}^{y}\left[\prod_{n<l<m}(-\sigma_{l}^{z})\right]\sigma_{m\in w}^{x} & = & i\eta_{n}\eta_{m},
\end{eqnarray}
\end{subequations}
where local links $(n,m)$ with $m=n+\hat{1}$ give rise to two-spin interactions and nonlocal links $(n,m)$ give rise to multiple-spin interactions. 
The additional two-spin and multiple-spin defined on the link $(n,m)$ should obey the following {\em supplementary rule}, 
\begin{enumerate}
\item $n$ and $m$ are not allowed to coincide with sites connected by existing $z$-bonds (but not other types of bonds) in the original Hamiltonian constructed subjet to two elementary rules.
\end{enumerate}
So that these additional spin interactions commute with existing $\hat{D}_{nm}$ in the original Hamiltonian and would not spoil the exact solvability and are able to lift the local degeneracy.

{\em Duality.} If one interchanges $w$ with $b$ and vice versa in the above, the fermionized $\eta$ Majorana fermion terms in Eqs.~\eqref{eq:eta} will change to $\gamma$ Majorana fermion terms in Eqs.~\eqref{eq:gamma}. 
Thus Eqs.~\eqref{eq:gamma} is dual to Eqs.~\eqref{eq:eta}, and {\em there exists a duality between $\eta$ and $\gamma$ Majorana fermions}.
Note that there is a similar duality symmetry relating topologically trivial and nontrivial phases in the interacting Kitaev chains\cite{Miao}.

{\em Shortcut multiple-spin interactions.} To couple Majorana fermions of the same species, say, $\eta$ or $\gamma$, on a nonlocal link $(n,m)$, we introduce multiple-spin interactions in the above. 
As examined in the original Kitaev honeycomb model, multiple-spin interactions can be added to the Hamiltonian without spoiling the exact solvability,
which to couple Majorana fermions of the same species on nonlocal link $(n,m)$ as well\cite{Kitaev,lee2007}. 
However, these multiple-spin interactions may contain infinite number of spin operators $\sigma_{l}^{z}$ in the thermodynamic limit, 
eventhough the spacial distance between site $n$ and $m$, $\sum_{j=1}^{d}\left|n_{j}-m_{j}\right|$, is small.
This will happen when the string spin operator $\prod_{n<l<m}\sigma_{l}^{z}$ winds around the system, say, $n_{j}\neq m_{j}$ for at least one $j\ge 2$. 
This is mathematically exact but hard to realize in a realistic physical system. 
Below we shall construct some {\em shortcut} multiple-spin interactions in addition to those in Eqs.~\eqref{eq:eta}, 
which consist of finite number of spin operators in the thermodynamic limit and remain the exact solvability as well.  

We begin with a concrete example and consider two sites $n\in w$ and $n+\hat{1}\in b$, and a $z$-bond on the link $(n+\hat{1},n+\hat{1}+\hat{2})$, where $\hat{j}$ is the unit vector along the $j$-th direction as defined after Eq.~\eqref{eq:Jzz1}.
The following multiple-spin interactions serves as one of the shortcut interactions,
\begin{equation}\label{eq:mul_ex}
i\sigma_{n}^{x}\sigma_{n+\hat{1}}^{x}\sigma_{n+\hat{1}}^{z}\sigma_{n+\hat{1}+\hat{2}}^{z}=\sigma_{n}^{x}\sigma_{n+\hat{1}}^{y}\sigma_{n+\hat{1}+\hat{2}}^{z},
\end{equation}
which couples the $\gamma$ Majorana fermions on the nonlocal link $(n,n+\hat{1}+\hat{2})$ as well as the $\eta$ Majorana fermions on the local link $(n+\hat{1},n+\hat{1}+\hat{2})$.
This can be seen by applying the Jordan-Wigner transformation. Then Eq.~\eqref{eq:mul_ex} becomes
\begin{eqnarray}
 & & \gamma_{n}\gamma_{n+\hat{1}}i\hat{D}_{n+\hat{1},n+\hat{1}+\hat{2}}\gamma_{n+\hat{1}}\gamma_{n+\hat{1}+\hat{2}}\nonumber \\
 &=& i\hat{D}_{n+\hat{1},n+\hat{1}+\hat{2}}\gamma_{n}\gamma_{n+\hat{1}+\hat{2}}
\end{eqnarray}
where the relation $\gamma^2_{n}=1$ is used. 
If there is an existing $z$-bond on the local link $(n+\hat{1},n+\hat{1}+\hat{2})$, this shortcut spin term will commute with all the $\hat{D}_{nm}$'s and guarantee the exact solvability. 

Now it is clear how to construct a specific shortcut multiple-spin interaction with the help of local bonds and existing $z$-bonds, 
and we shall present a generic way to construct a shortcut multiple-spin interaction which couples Majorana fermions of the same species on a nonlocal link $(n,m)$.
Thanks to the $\gamma\leftrightarrow \eta$ duality, we construct spin terms for $\gamma$ Majorana fermions only and those for the $\eta$ Majorana fermions can be constructed by the duality, say, switching $w$ and $b$ sublattices. 


To do this, we consider a path connecting site $n$ and $m$, which consists of finite number of local links, namely, links of the form $(l,l+\delta)$ with $\delta=\hat{j}$ defined after Eq.~\eqref{eq:Jzz1}.
Such a path is directional and we call each local link $(l,l+\delta)$ a {\em step}. 
A generic shortcut multiple-spin interaction can be constructed by assigning a two-spin terms on each step along the path and multiplying them together.
The path itself and the assignment of two-spin terms along the path are subject to the following {\em supplementary rules}:
\begin{enumerate}
\item For a step along the $\hat{1}$-direction, the two-spin term should be $\sigma_{l}^{\alpha}\sigma_{l+\hat{1}}^{\beta}$ with $\alpha,\beta=x,y$; 
for a step along the other directions, the two-spin terms should be $\sigma_{l}^{z}\sigma_{l+\delta}^{z}$ with $\delta\neq\hat{1}$, and there must exit a local $z$-bond on this step in the original Hamiltonian.
\item The indices $\alpha$ and $\beta$ should be chosen as follows: for $n\in w$ and $n+\hat{1}\in b$, $(\alpha,\beta)=(x,x)$;
for $n\in b$ and $n+\hat{1}\in w$, $(\alpha,\beta)=(y,y)$;
for $n\in w$ and $n+\hat{1}\in w$, $(\alpha,\beta)=(x,y)$;
for $n\in b$ and $n+\hat{1}\in b$, $(\alpha,\beta)=(y,x)$.
\end{enumerate}
After the Jordan-Wigner transformation, such a shortcut multiple-spin interaction reads
\begin{equation}
\left(\prod_{l\in path,\delta\neq\hat{1}} \hat{D}_{l,l+\delta}\right)i\gamma_{n}\gamma_{m},
\end{equation}
where the product runs over allowed $z$-bonds along the path. 
As mentioned, to obtain a shortcut multiple-spin interaction coupling $\eta$ Majorana fermions on a nonlocal link $(n,m)$, what we need is to switch $w$ and $b$ to obtain dual terms 
from those corresponding to $\gamma$ Majorana fermions.

These shortcut multiple-spin interactions can be transformed to quadratic Majorana fermions coupled to a $Z_2$ background gauge field, it can be exactly diagonalized in each $\{D_{nm}\}$ sector as well.
Before the ending of this section, we would like to point out that such multiple-spin interactions can be generated perturbatively in the presence of an external magnetic field\cite{Kitaev}.

\section{Model classification}

\begin{figure}[hptb]
\begin{center}
\includegraphics[width=8cm]{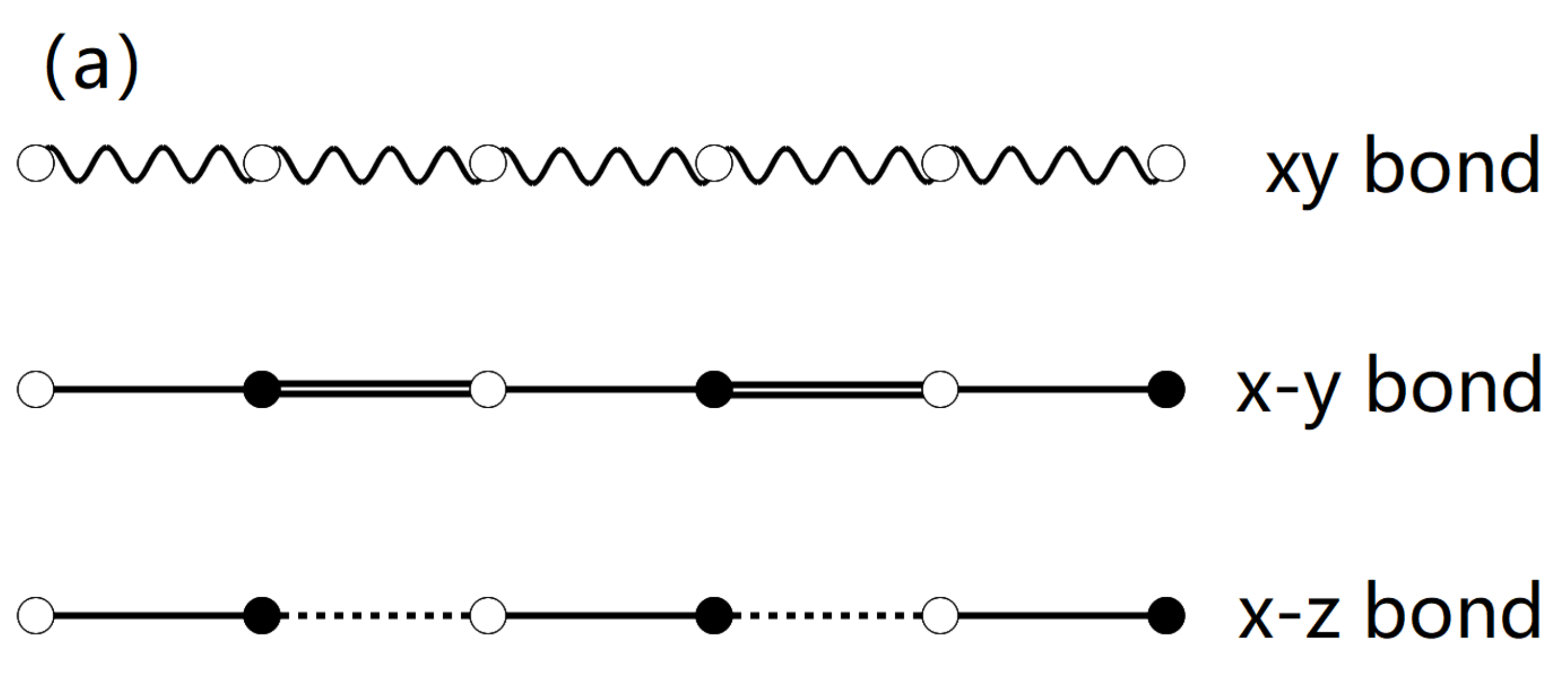}
\includegraphics[width=8.2cm]{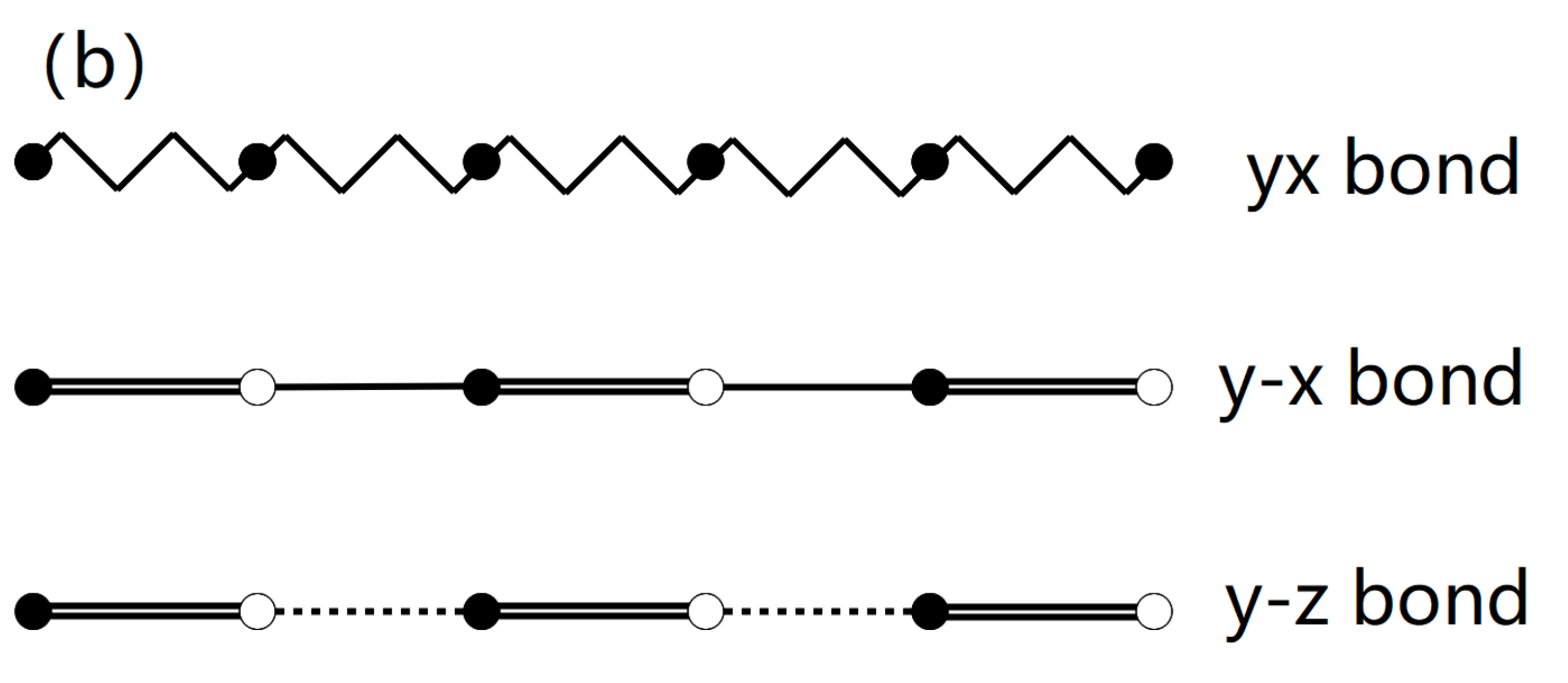}

\end{center}
\caption{Three parent spin models in $1D$ and their dual models. 
Solid lines denote local $x$-bonds, double solid lines denote local $y$-bonds, dashed lines denote local $z$-bonds, wavy lines denote local $xy$-bonds, and zigzag lines denote local $yz$-bonds. 
(a) Three parent models: $xy$ bond, $x$-$y$ bond, and $x$-$z$ bond chain model. 
(b) Three dual models to the parent models: $yx$ bond, $y$-$x$ bond, and $y$-$z$ bond chain model.}
\label{fig:1d}
\end{figure}

In this section, we shall classify the models constructed in previous sections according to their parent models, and carry out the classification in one, two, and three dimensions respectively.
The lattice translational symmetry is not necessary to the model construction and the exact solvability.
Nevertheless, for simplicity, we shall classify the models with translational symmetry only, and discuss disordered situation at the end of this section.  

{\em $1D$ spin models.} Even though the lattice structure is trivial, the class of $1D$ exactly solvable generalized Kitaev spin-$1/2$ models shares some universal properties with those in $D>1$. 
There are three parent spin models in $1D$ which are represented in FIG.~\ref{fig:1d}(a), namely, $xy$ bond, $x$-$y$ bond and $x$-$z$ bond chain model.
These parent models consist of local bonds only and are named according to the bonds in a unit cell.
A series of exactly solvable spin models can be generated from these three parent models by three operations and their combination as follows:

(i) Firstly, one can change all the white sites to black sites and vice versa. Then a new model can be generated according to the construction rules.
Actually, this operation is nothing but the duality discussed in previous section, and can be implemented through interchanging $\sigma^x$ with $\sigma^y$.
Note that the interchanging  $\sigma_{n}^x \to\sigma_{n}^y, \sigma_{n}^y\to\sigma_{n}^x$ is not a unitary transformation such that the $xy$ bond chain model is not equivalent to the $x$-$y$ bond chain model,
and the duality is not a unitary transformation in general.
For instance, one can generate three dual models from the three parent spin models by the duality operation, 
i.e. $yx$ bond, $y$-$x$ bond, and $y$-$z$ bond chain model as shown in FIG.~\ref{fig:1d}(b). 
This operation can always be implemented and hence each model has its dual model through the duality.

\begin{figure}[hptb]
\begin{center}
\includegraphics[width=9cm]{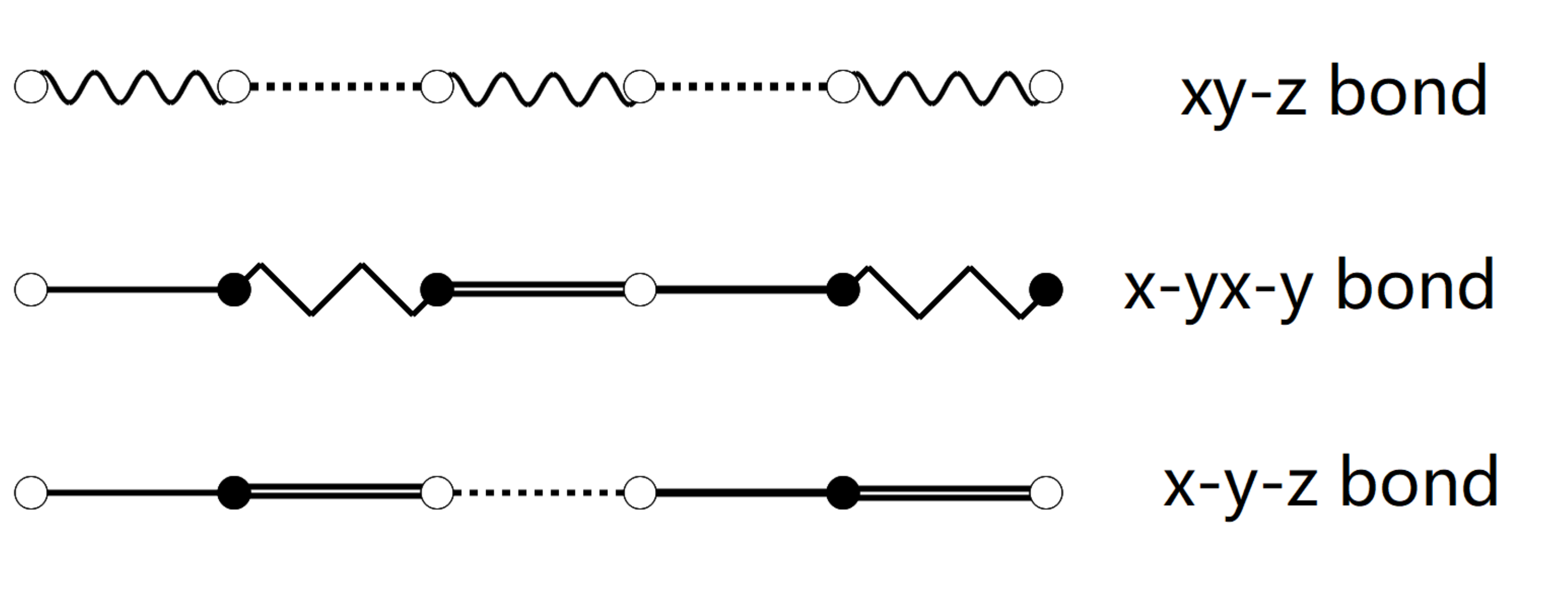}

\end{center}
\caption{Three spin models generated from $1D$ parent models by inserting local $z$-bonds, $yx$-bonds and $y$-bonds respectively:
$xy$-$z$ bond, $x$-$yx$-$y$ bond, and $x$-$y$-$z$ bond chain model.
Solid lines denote local $x$-bonds, double solid lines denote local $y$-bonds, dashed lines denote local $z$-bonds, wavy lines denote local $xy$-bonds, and zigzag lines denote local $yz$-bonds. 
}
\label{fig:1d_insert}
\end{figure}

(ii) Secondly, one can split one site into two and insert a local bond between these two sites subject to the construction rules.
The inserted local bond can be one of $x$-bond, $y$-bond, $z$-bond, $xy$-bond and $yx$-bond, and the construction rules should be respected.
For instance, three new spin models can be constructed by splitting one site and inserting $z$-bonds, $yx$-bonds and $y$-bonds to the three parent models respectively,
as shown in FIG.~\ref{fig:1d_insert}. We call them $xy$-$z$ bond, $x$-$yx$-$y$ bond, and $x$-$y$-$z$ bond chain model, which follows the arrangement of local bonds in the enlarged unit cell.


\begin{figure}[hptb]
\begin{center}
\includegraphics[width=8cm]{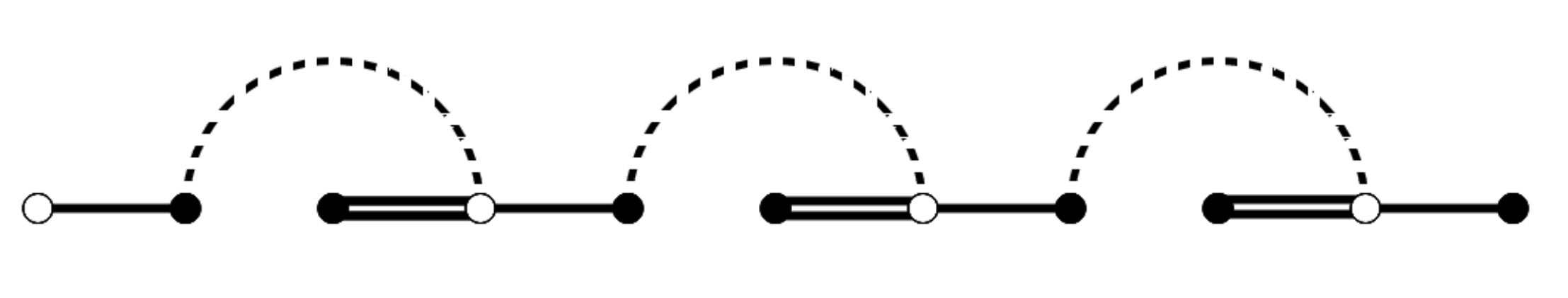}

\end{center}
\caption{The spin model generated from $x$-$yx$-$y$ bond chain model.
Solid lines denote local $x$-bonds, double solid lines denote local $y$-bonds, dashed lines denote non-local $z$-bonds.
}
\label{fig:1d_erase}
\end{figure}

(iii) Thirdly, one can erase existing local bonds, and/or add nonlocal bonds subjecting to construction rules. This operation does not add or remove any site.
The 1D chain may become two disconnected chains by erasing an existing local bond. However, the added nonlocal bonds can rescue this as shown in FIG.~\ref{fig:1d_erase}, which model is generated from $x$-$yx$-$y$ bond chain model by erasing the $yx$-bonds and add nonlocal $z$-bonds.

In principle, we can repeat these operations and generate infinite numbers of exactly solvable spin models from the three parent models. 


{\em $2D$ spin models.} According to the construction rules, a $2D$ exactly solvable models can be constructed by coupling the $1D$ models through $z$-bonds only.
Note that a shortcut multiple-spin interaction depends on existing $z$-bonds, which can be added after the $2D$ model is constructed.
As an example, a $2D$ spin model can be constructed by coupling $x$-$y$ bond chain and $y$-$x$ bond chain (see FIG.~\ref{fig:1d}) alternatively through $z$-bonds 
as illustrated in FIG.~\ref{fig:honey}(a), and which is topologically equivalent to a honeycomb model plotted in FIG.~\ref{fig:honey}(b). 
This $2D$ model is nothing but the original Kitaev honeycomb model\cite{Kitaev} in the brick wall representation\cite{Feng}. 
It turns out that there are only two parent spin models in $2D$. One is the Kitaev honeycomb model and the other is $xy$ bond honeycomb model as shown in FIG.~\ref{fig:honey}(c) and (d). 
Similar to $1D$, a series of exactly solvable models can be generated starting from these parent models by the following operations:

\begin{figure}[hptb]
\begin{center}
\includegraphics[width=4.5cm]{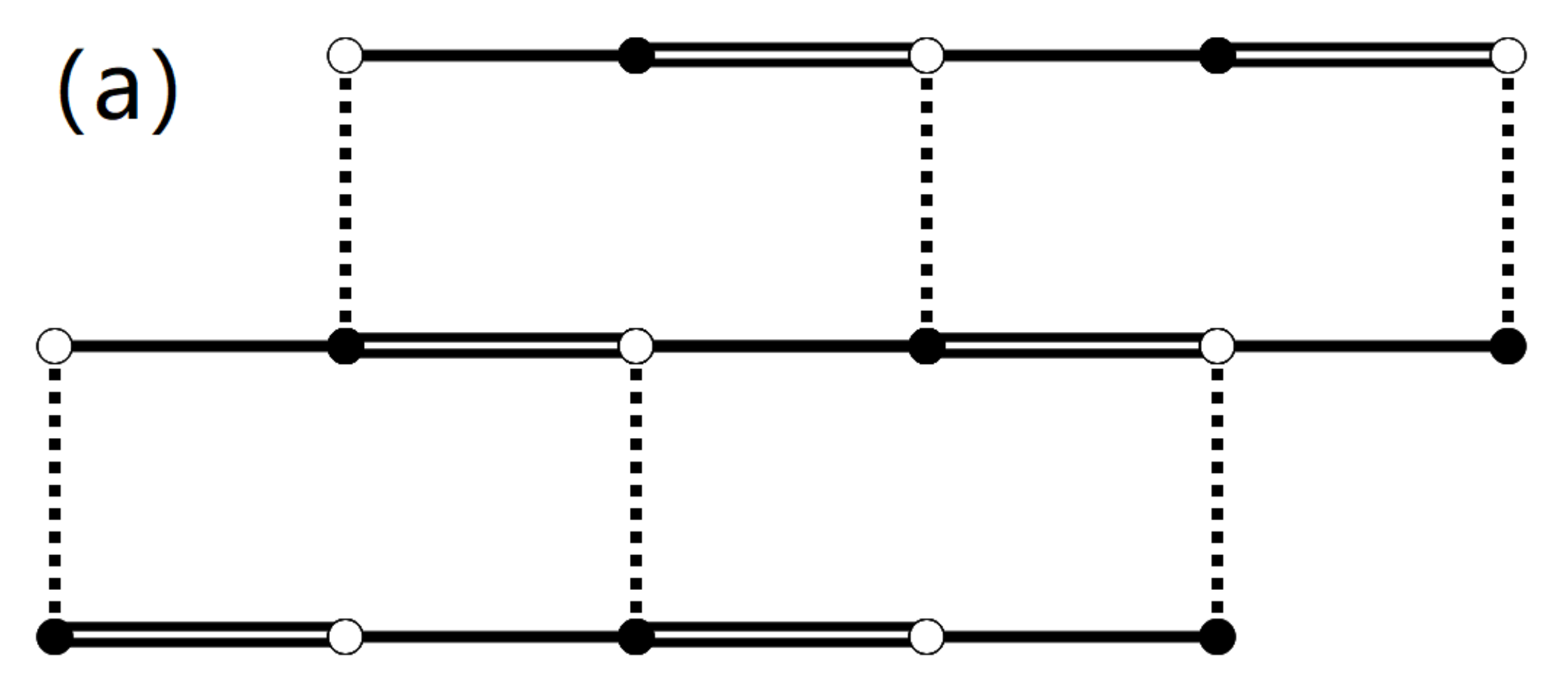}
\includegraphics[width=3.5cm]{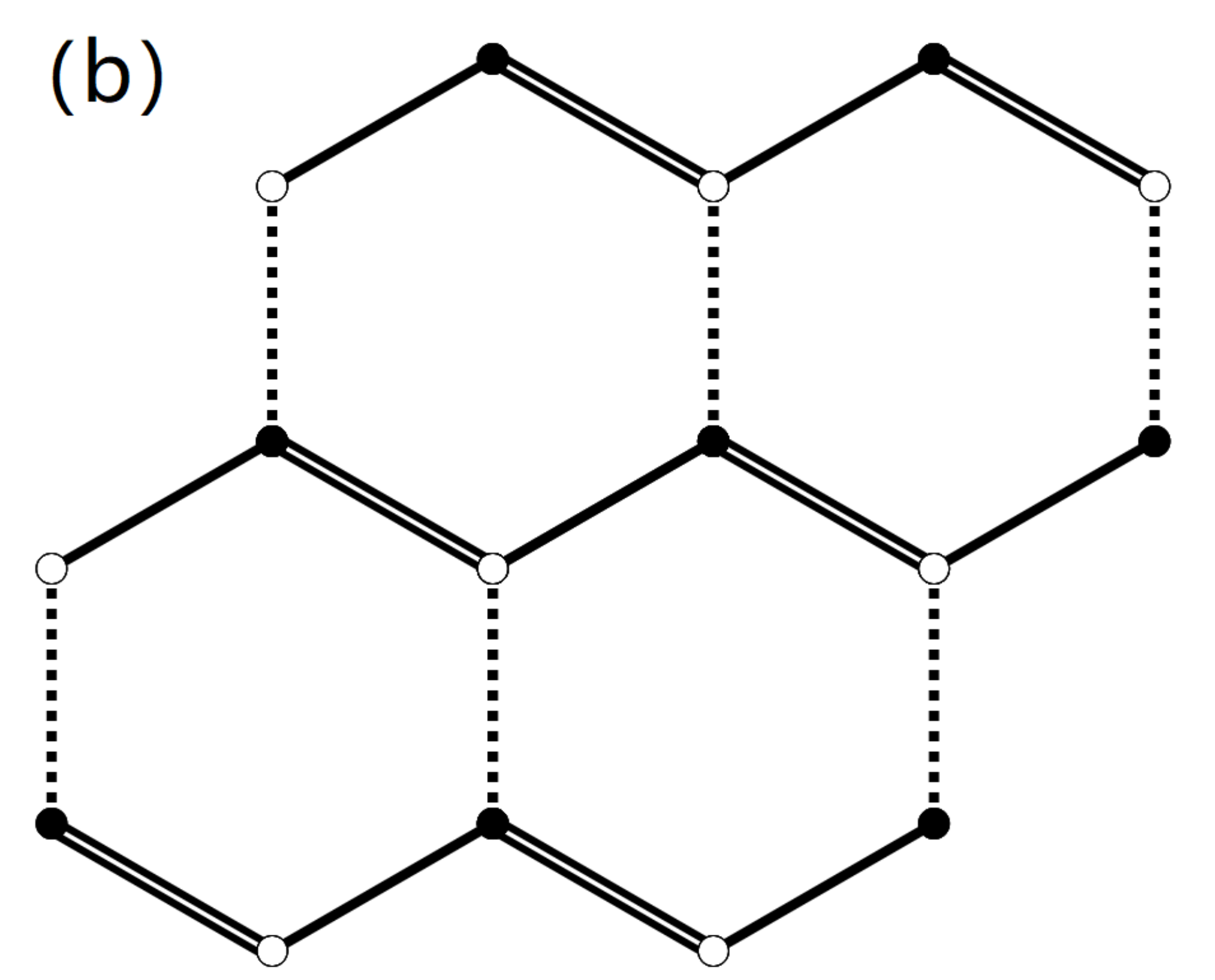}
\includegraphics[width=4.5cm]{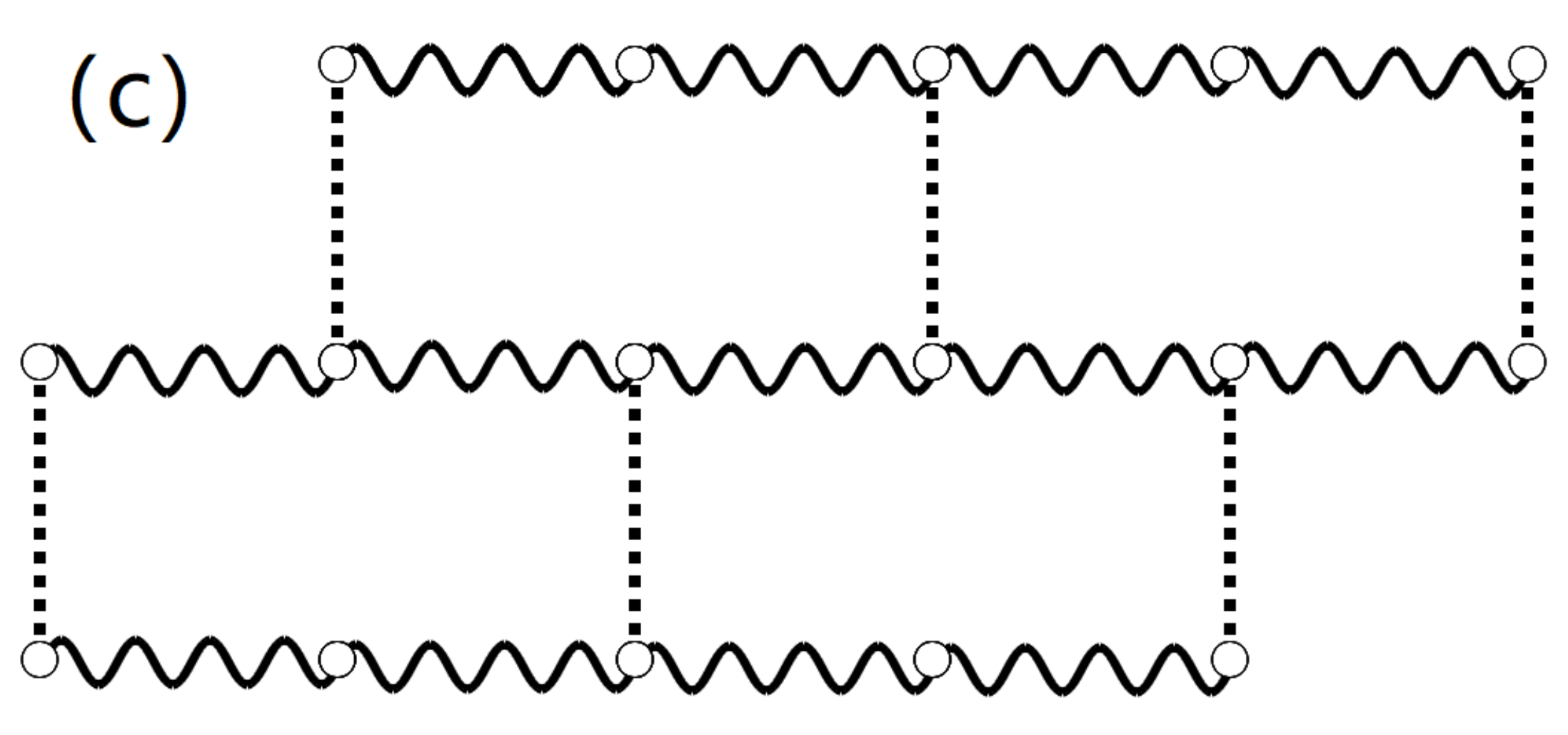}
\includegraphics[width=3.5cm]{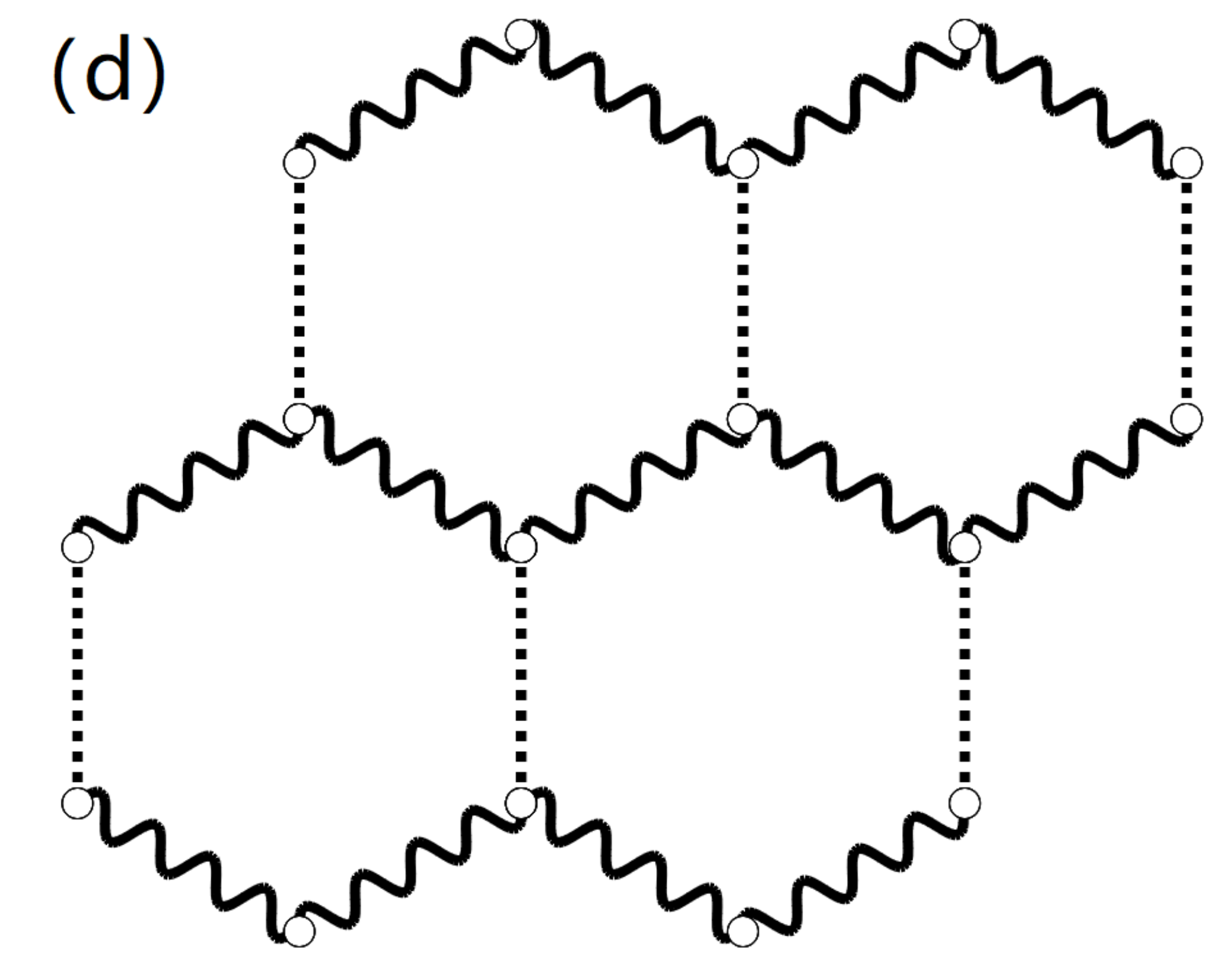}
\end{center}
\caption{Two parent models in $2D$: Kitaev honeycomb model and $xy$ bond honeycomb model. 
Solid lines denote local $x$-bonds, double solid lines denote local $y$-bonds, dashed lines denote local $z$-bonds, and wavy lines denote local $xy$-bonds. 
(a) Kitaev honeycomb model in brick wall representation. (b) Kitaev honeycomb model. (c) $xy$ bond honeycomb model in brick wall representation. (d) $xy$ bond honeycomb model.}
\label{fig:honey}
\end{figure}

\begin{figure}[hptb]
\begin{center}
\includegraphics[width=4.5cm]{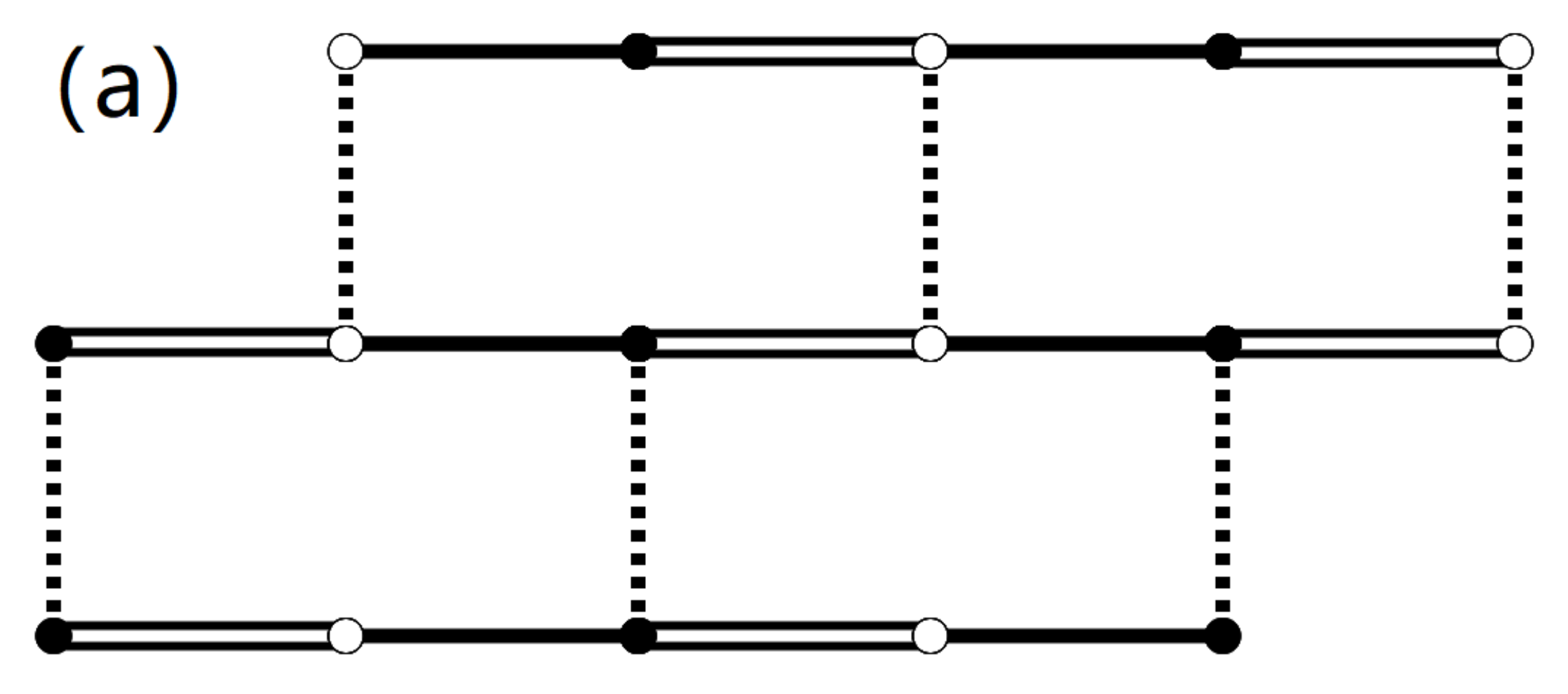}
\includegraphics[width=3.5cm]{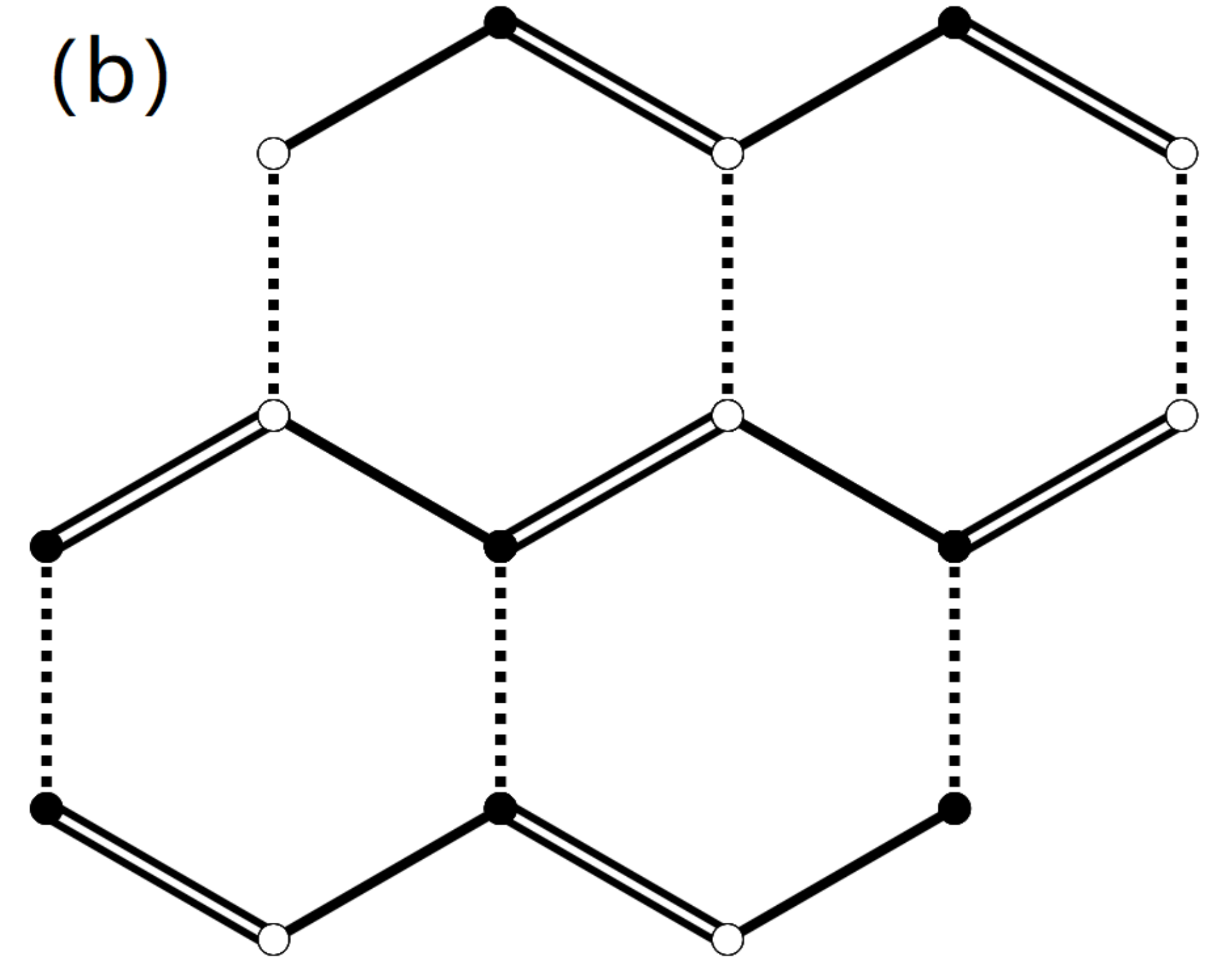}
\end{center}
\caption{One derivant model from Kitaev honeycomb model and its topologically equivalent brick wall representation. 
Solid lines denote local $x$-bonds, double solid lines denote local $y$-bonds, and dashed lines denote local $z$-bonds. 
(a) The derivation model from Kitaev honeycomb model in brick wall representation. (b) The derivation model from Kitaev honeycomb model.}
\label{fig:counter}
\end{figure}

(i) Firstly, one can perform the duality transformation along a single chain, namely, switch white and black sites and re-allocate local bonds according to the construction rules.
Note that it is different from the $1D$ case where each model has only one dual model, a $2D$ spin model has $2^{L_2}-1$ derivant models in a system consisting of $L_2$ chains, 
since the operation changes each chain independently. If one performs the duality transformation in every chains, the corresponding derivant model is called the dual model.
Moreover, these operations may change the number of sites per unit cell in $2D$ and even give rise to randomness along the $\hat{2}$-direction. 
One derivant model from Kitaev honeycomb model is constructed by performing the duality transformation along all the even number-th chains as demonstrated in FIG.~\ref{fig:counter}, 
which possesses four sites per unit cell in comparison with two sites per unit cell in the original Kitaev honeycomb model.

(ii) Secondly, similar to the $1D$ case, one can split sites and insert local bonds (say, $x$-bond, $y$-bond, $z$-bond, $xy$-bond and $yx$-bond) and/or add nonlocal bonds directly to the parent spin models according to the construction rules.
For instance, a chiral spin liquid model defined on triangle honeycomb lattice\cite{yao2007} can be generated from the Kitaev honeycomb model by inserting local $x$-bonds and $y$-bonds and adding nonlocal $z$-bonds.

\begin{figure}[hptb]
\begin{center}
\includegraphics[width=8cm]{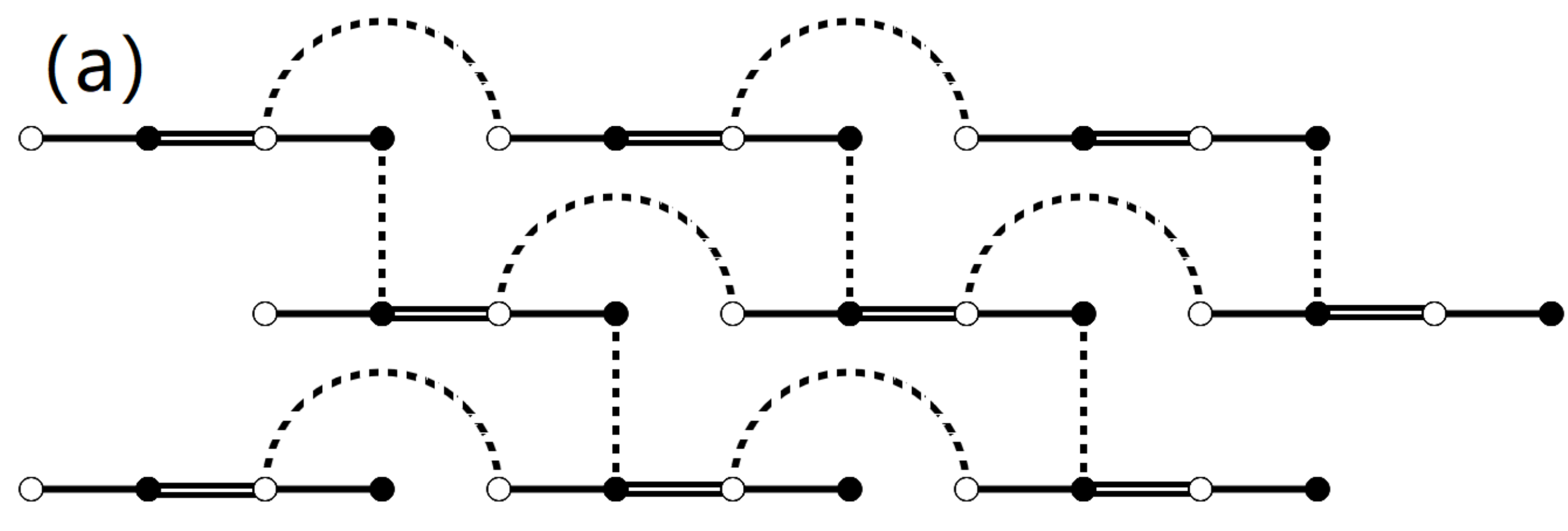}
\includegraphics[width=5cm]{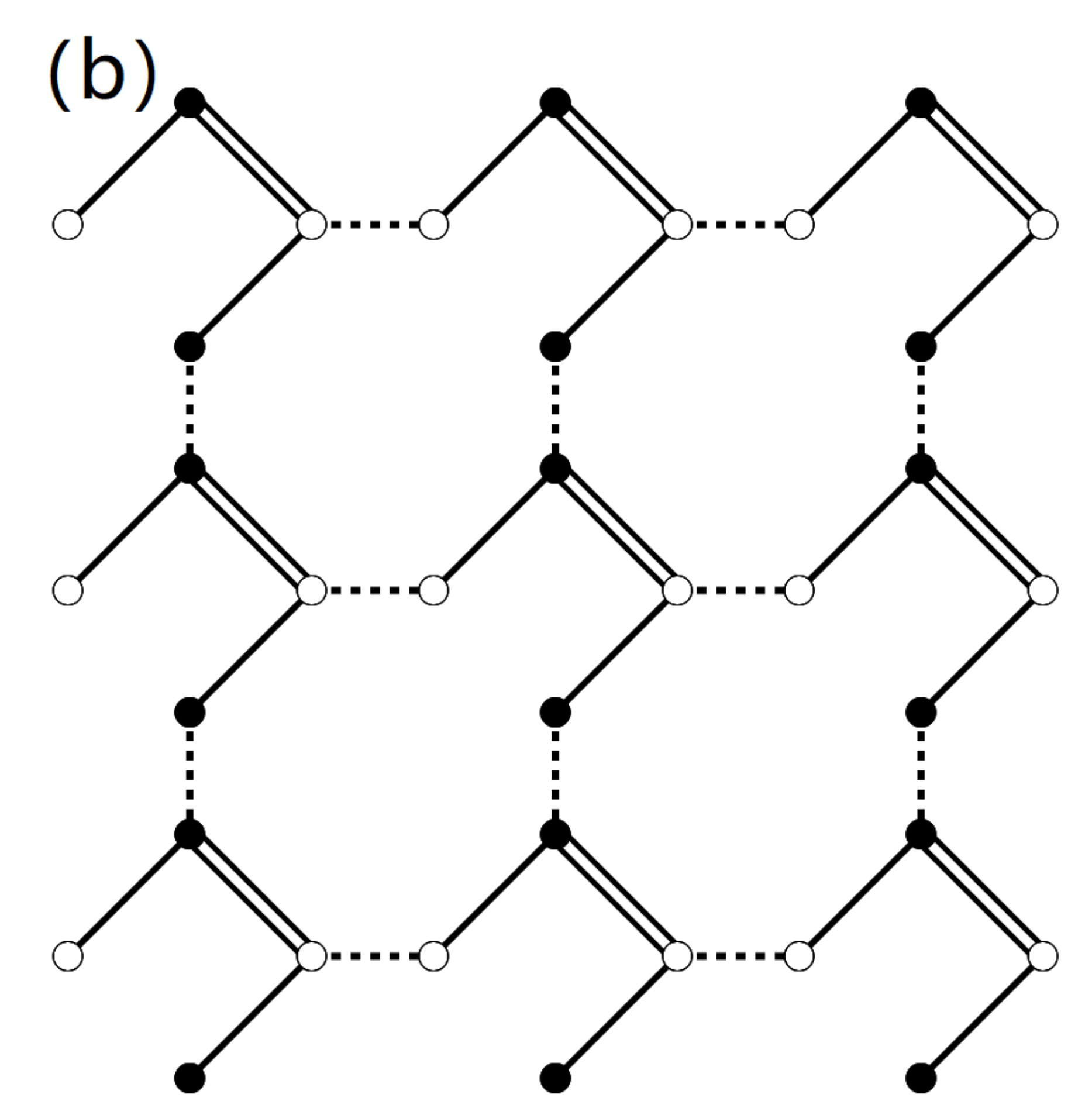}
\end{center}
\caption{Square-octagon model and its topologically equivalent brick wall representation. 
Solid lines denote local $x$-bonds, double solid lines denote local $y$-bonds, and dashed lines denote $z$-bonds. 
(a) Square-octagon model in brick wall representation. (b) Square-octagon model.}
\label{fig:squ_oct}
\end{figure}

So far, for all the models that we enumerate in $2D$, the local bonds form a connected graph on the lattice. Nevertheless, it is allowed to construct a model in which local bonds constitute disconnected clusters only.
A concrete model is constructed as shown in FIG.~\ref{fig:squ_oct}, which also define the square-octagon lattice. 
It can be seen from its topologically equivalent brick wall representation that the nonlocal $z$-bonds are crucial to construct such a $2D$ lattice. 
Thus similar to the $1D$ case, we have the following operation to generate new spin models:

(iii) Thirdly, one can erase a local bond (or leave it empty). If the remaining bonds do not form a connected graph, we can add nonlocal bonds, which are subject to the construction rules, to connect the sites originally belonging to the erased bond.
The square-octagon spin model in FIG.~\ref{fig:squ_oct} can be generated from the Kitaev honeycomb model by the combination of these three types of operations. 


\begin{figure}[hptb]
\begin{center}
\includegraphics[width=8cm]{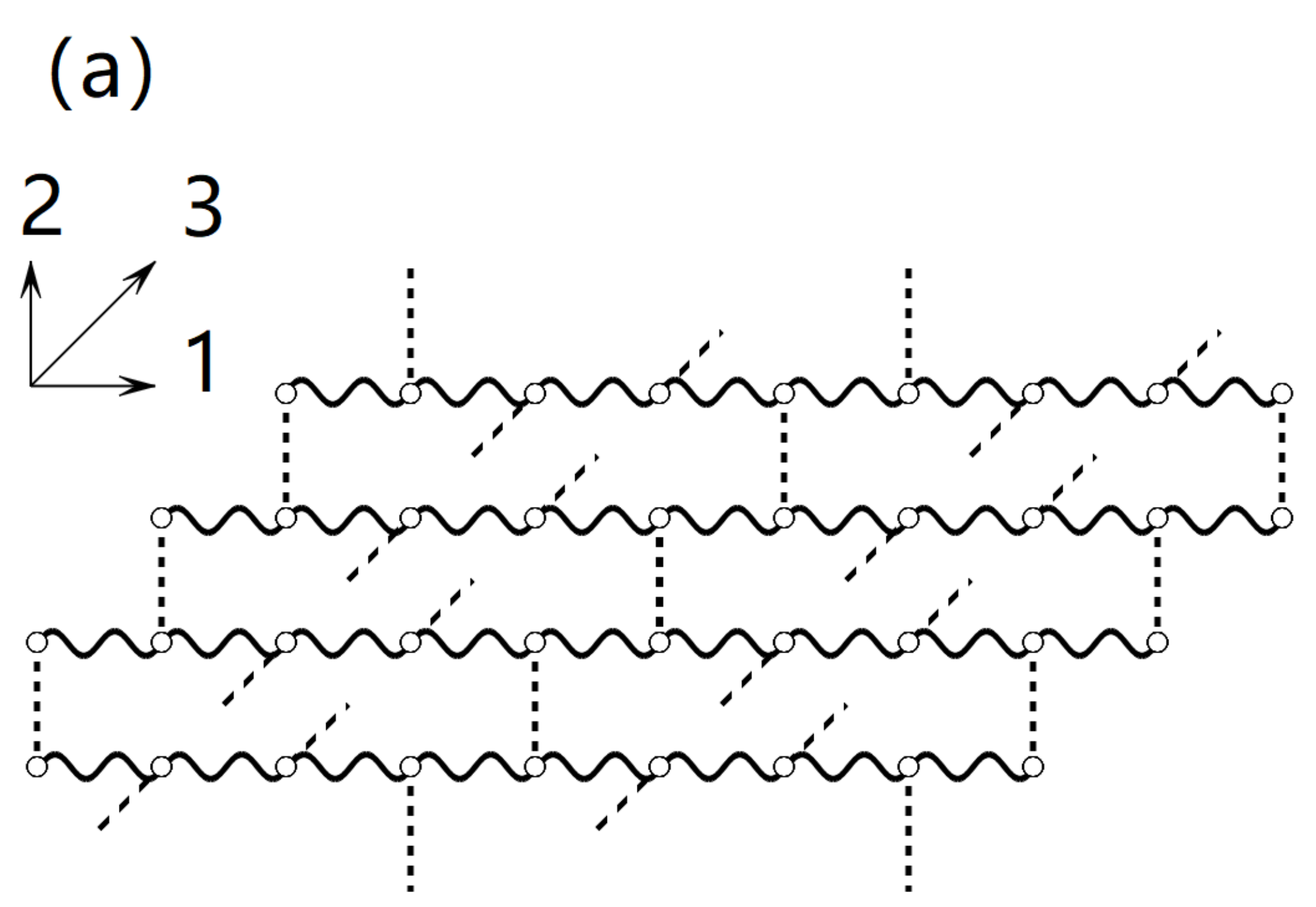}
\includegraphics[width=8cm]{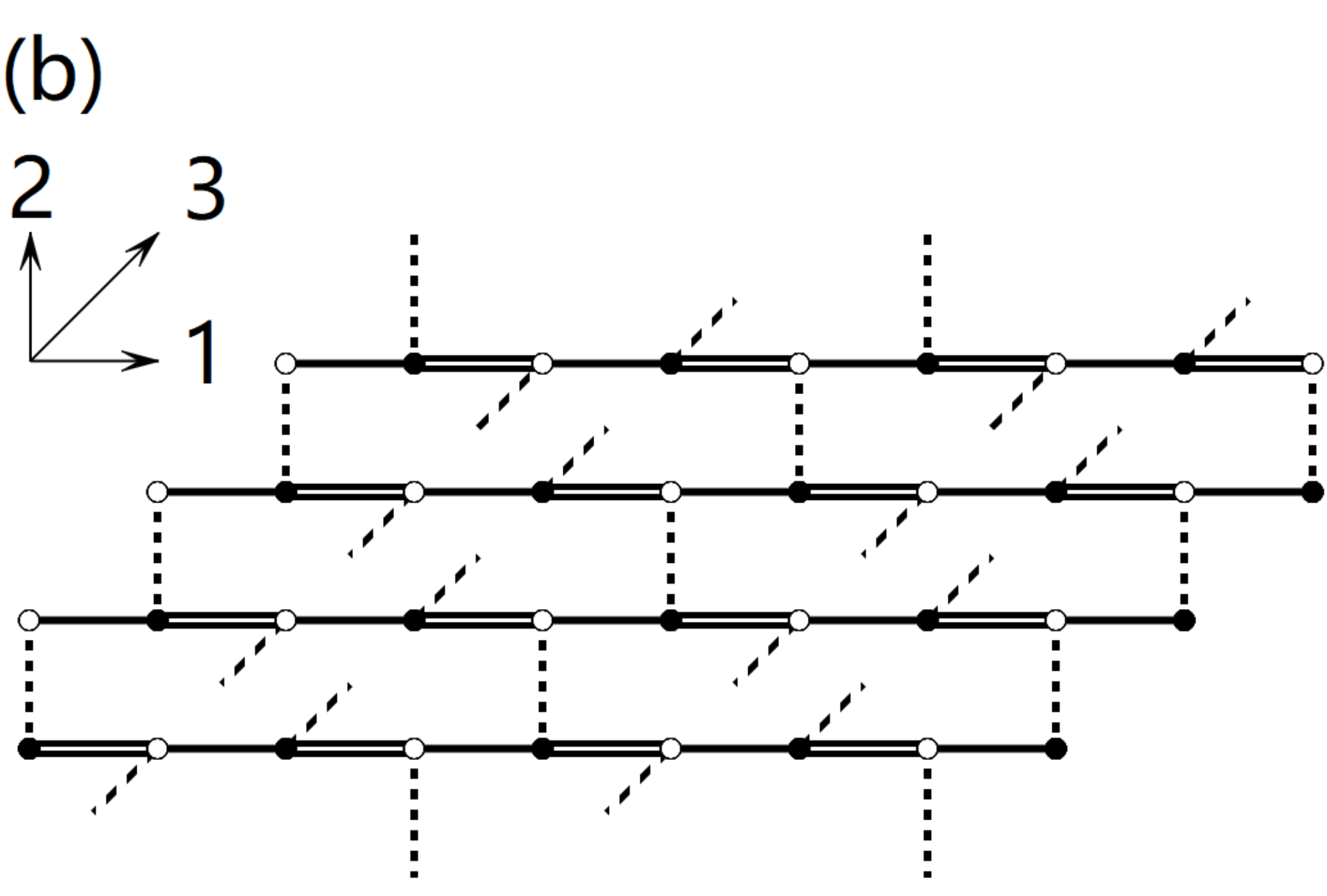}
\end{center}
\caption{Two parent models in $3D$: (a) $3D$ $xy$ bond model and (b) $3D$ $x$-$y$ bond model. 
Solid lines denote local $x$-bonds, double solid lines denote local $y$-bonds, vertical (horizontal) dashed lines denote local $z$-bonds in the $\hat{2}$-direction and $\hat{3}$-direction,
wavy lines denote local $xy$-bonds.}
\label{fig:3d}
\end{figure}

\begin{figure}[hptb]
\begin{center}
\includegraphics[width=6cm]{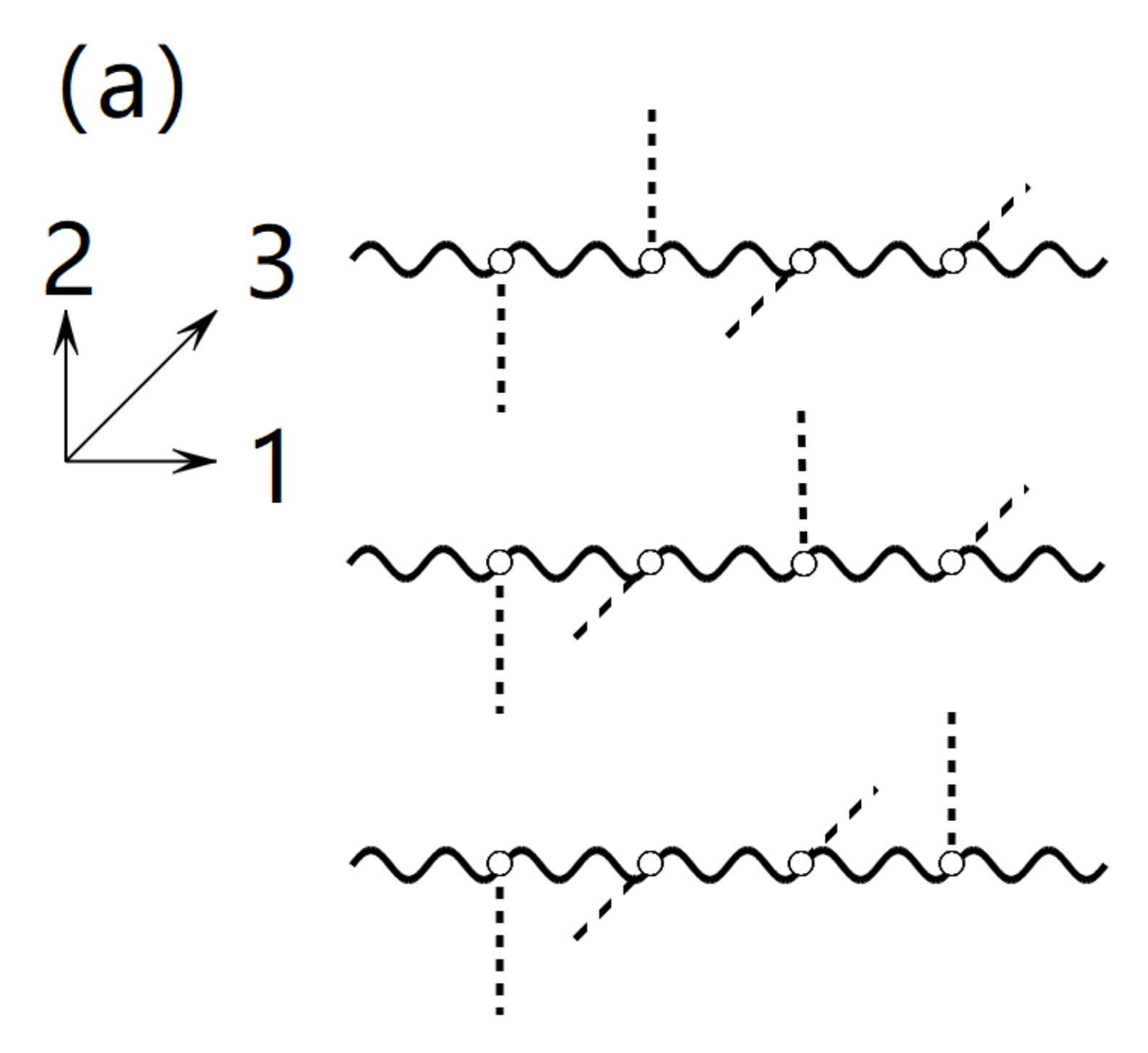}
\includegraphics[width=6cm]{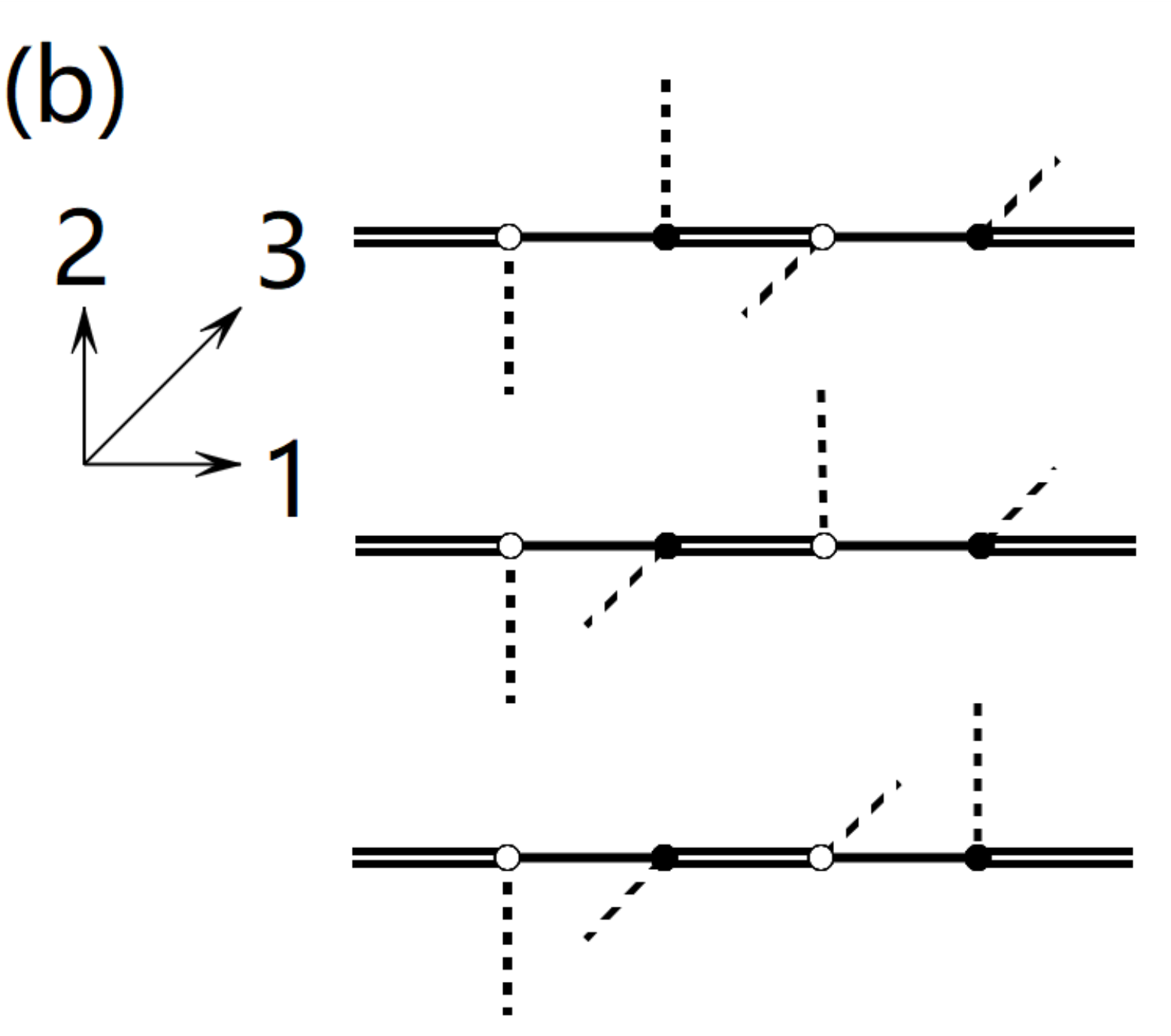}
\end{center}
\caption{Three types of unit cells for (a) $3D$ $xy$ bond model and (b) $3D$ $x$-$y$ bond model. 
Solid lines denote local $x$-bonds, double solid lines denote local $y$-bonds, vertical (inclined) dashed lines denote local $z$-bonds in the $\hat{2}$-direction ($\hat{3}$-direction),
wavy lines denote local $xy$-bonds.}
\label{fig:3d_unit}
\end{figure}

{\em $3D$ spin models.} A $3D$ exactly solvable generalized Kitaev spin model can be constructed by coupling $1D$ chains along the $\hat{2}$- and $\hat{3}$-directions using $z$-bonds only according to the construction rules. 
There are two parent spin models in $3D$ as shown in FIG.~\ref{fig:3d}, which are called $3D$ $xy$ bond model and $3D$ $x$-$y$ bond model respectively. 
Note there are three types of unit cell for each model distinguished by the ordering of local $z$-bonds as shown in FIG.~\ref{fig:3d_unit}. 
Regarding the ordering of these ``glue" $z$-bonds and the single chain, we have $3\times 2$ subclasses of models indeed. 
One can generate a series of new exactly solvable models from one of these parent spin models by the same operations as in $2D$. 

Before the end of this section, we would like to make the following remarks:
(1) Our classification for these exactly solvable generalized Kitaev spin models through their parent models are heuristic. 
A rigorous classification requires the knowledge of group theory and we leave it for future study.
(2) Translational symmetry or periodicity is not necessary to the exact solvability. We can construct non-periodic exactly solvable spin models as long as the construction rules are respected. 
There are two sources of non-periodicity. One comes from the non-periodic distribution of bonds. The other comes from the spatial dependent coupling constants even if the bonds are allocated periodically. 
Neither of them spoils the exact solvability.

\section{Examples}
In this section, we demostrate how to solve the generalized Kitaev spin models exactly through two exmaples: a $2D$ tetragon-octagon model and a $3D$ $xy$ bond model.
Other models can be solved in the same strategy.

\begin{figure}[hptb]
\begin{center}
\includegraphics[width=6cm]{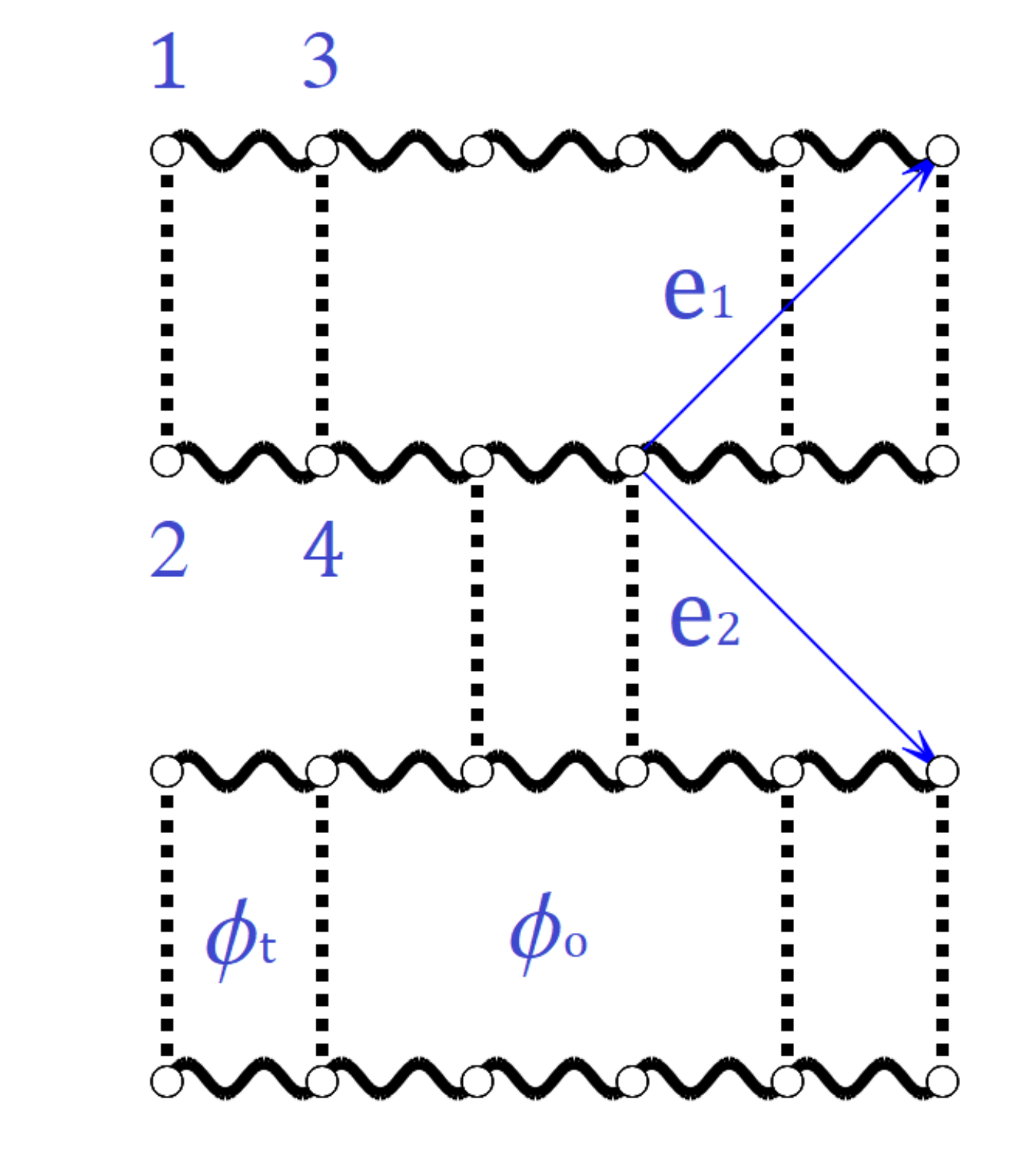}
\end{center}
\caption{Tetragon-octagon lattice. Wave lines denote local $xy$-bonds, and dashed lines denote local $z$-bonds. 
$e_{1}=\left(1,1\right)$ and $e_{2}=\left(1,-1\right)$ are the two basis. Each tetragon plaquette contains four sites labeled as $\mu=1,2,3,4$.
$\phi_t$ and $\phi_o$ are $Z_2$ flux operators defined on tetragon and octagon plaquettes and by Eqs.\eqref{eq:phit} and \eqref{eq:phio} respectively.
}
\label{fig:tet_oct}
\end{figure}

{\em Example I: $2D$ tetragon-octagon model.} The lattice of tetragon-octagon model is shown in FIG.~\ref{fig:tet_oct}.
It is topologically equivalent to the lattice of the 4-8-8 mosaic model studied in Ref.\cite{yang}. 
But these two models are different from each other on the spin interactions. 
The the 4-8-8 mosaic model consists of three types of local bonds, $x$-bonds, $y$-bonds and $z$ bonds, while our tetragon-octagon model consists of only two types of local bonds, $xy$-bonds and $z$-bonds.
The Hamiltonian of the tetragon-octagon model reads,
\begin{eqnarray}
H & = & \sum_{\vec{r}}J_{1}^{xy}\sigma_{\vec{r},1}^{x}\sigma_{\vec{r},3}^{y}+J_{1}^{xy}\sigma_{\vec{r},2}^{x}\sigma_{\vec{r},4}^{y}\nonumber \\
  &   & +J_{2}^{xy}\sigma_{\vec{r},3}^{x}\sigma_{\vec{r}+e_{1},2}^{y}+J_{2}^{xy}\sigma_{\vec{r},4}^{x}\sigma_{\vec{r}+e_{2},1}^{y}\nonumber \\
  &   & +J^{z}\sigma_{\vec{r},1}^{z}\sigma_{\vec{r},2}^{z}+J^{z}\sigma_{\vec{r},3}^{z}\sigma_{\vec{r},4}^{z},
\end{eqnarray}
where the unit cell is chosen as a tetragon plaquette and denoted by a Bravais vector $\vec{r}=(r_1,r_2)$, a lattice site is then labeled as $\left(\vec{r},\mu\right)$ with sublattice indices $\mu=1,2,3,4$. 
The two Bravais lattice basis vectors are $e_{1}=\left(1,1\right)$ and $e_{2}=\left(1,-1\right)$  as shown in FIG.~\ref{fig:tet_oct}. 
$J_{1}^{xy}$ and $J_{2}^{xy}$ are intra-unit-cell and inter-unit-cell $xy$-bond  coupling constants respectively. 
$J^{z}$ is the $z$-bond coupling constant, which is an intra-unit-cell coupling.
The fermionization through the Jordan-Wigner transformation gives rise to
\begin{eqnarray}
H & = & \sum_{\vec{r}}-iJ_{1}^{xy}\gamma_{\vec{r},1}\gamma_{\vec{r},3}-iJ_{1}^{xy}\gamma_{\vec{r},2}\gamma_{\vec{r},4}\nonumber \\
  &   & -iJ_{2}^{xy}\gamma_{\vec{r},3}\gamma_{\vec{r}+e_{1},2}-iJ_{2}^{xy}\gamma_{\vec{r},4}\gamma_{\vec{r}+e_{2},1}\nonumber \\
  &   & +iJ^{z}\hat{D}_{\vec{r},1}\gamma_{\vec{r},1}\gamma_{\vec{r},2}+iJ^{z}\hat{D}_{\vec{r},3}\gamma_{\vec{r},3}\gamma_{\vec{r},4},
\end{eqnarray}
where $\hat{D}_{\vec{r},\mu}=-i\eta_{\vec{r},\mu}\eta_{\vec{r},\mu+1}$ and $\hat{D}_{\vec{r},\mu}^2=1$.

For open boundary condition, all the $\hat{D}_{\vec{r},\mu}$'s commute with each other and with the Hamiltonian. Hence all $\hat{D}_{\vec{r},\mu}$ are constants of motion and can be replaced by their eigenvalues $\pm1$.
To determine the ground state(s), we need to identify the sector(s), $\{D_{\vec{r},\mu}\}$, minimizing the total energy. Note the tetragon-octagon lattice contains two types of plaquettes, tetragon and octagon plaquettes. 
We define the $Z_2$ flux operator on the tetragon plaquette as 
\begin{subequations}
\begin{eqnarray}
\phi_{t} & = & -\sigma_{\vec{r},1}^{y}\sigma_{\vec{r},3}^{x}\sigma_{\vec{r},4}^{x}\sigma_{\vec{r},2}^{y} \nonumber\\
         & = & \hat{D}_{\vec{r},1}\hat{D}_{\vec{r},3}, \label{eq:phit}
\end{eqnarray}
and its counterpart on the octagon plaquette as
\begin{eqnarray}
\phi_{o} & = & -\sigma_{\vec{r},3}^{y}\sigma_{\vec{r}+e_{1},2}^{z}\sigma_{\vec{r}+e_{1},4}^{z}\sigma_{\vec{r}+e_{1}+e_{2},1}^{x} \nonumber \\
         &   & \times\sigma_{\vec{r}+e_{1}+e_{2},2}^{x}\sigma_{\vec{r}+e_{2},3}^{z}\sigma_{\vec{r}+e_{2},1}^{z}\sigma_{\vec{r},4}^{y} \nonumber\\
         & = & \hat{D}_{\vec{r},3}\hat{D}_{\vec{r}+e_{1}+e_{2},1}. \label{eq:phio}
\end{eqnarray}
\end{subequations}
It is easy to verify that all these $Z_2$ flux operators commute with each other and with the Hamiltonian, and $\hat{\phi}_{t}^{2}=\hat{\phi}_{o}^{2}=1$. 
So we can replace each $\hat{\phi}_{t}$ and $\hat{\phi}_{o}$ by their eigenvalues $\phi_t=\pm1$ and $\phi_o=\pm1$. 
Numerically, we find that the ground states are all $\pi$-flux states, i.e., $\phi_{t}=\phi_{o}=-1$ everywhere, 
which is remarkably different from the Kitaev honeycomb model, whose ground states are zero flux states. 
Thus the ground state degeneracy of the tetragon-octagon model is of $2^{L_2-1}$-fold under the open boundary condition. 
These degenerate ground states are given by all the possible $\{D_{\vec{r},\mu}\}$ giving rise to $\pi$-fluxes on every tetragon and octagon plaquettes.

For periodic boundary condition, additional boundary terms will emerge from the Jordan-Wigner transformation\cite{yao2007},
\begin{align}
 & J_{2}^{xy}\sigma_{\left(L_{1},r_{2}\right),3}^{x}\sigma_{\left(1,r_{2}+1\right),2}^{y}+J_{2}^{xy}\sigma_{\left(L_{1},r_{2}\right),4}^{x}\sigma_{\left(1,r_{2}-1\right),1}^{y}\nonumber \\
= & -iJ_{2}^{xy}\hat{F}_{r_{2}}^{+}\gamma_{\left(L_{1},r_{2}\right),3}\gamma_{\left(1,r_{2}+1\right),2}\nonumber \\
 & -iJ_{2}^{xy}\hat{F}_{r_{2}}^{-}\gamma_{\left(L_{1},r_{2}\right),4}\gamma_{\left(1,r_{2}-1\right),1},
\end{align}
with
\begin{subequations}
\begin{align}
\hat{F}_{r_{2}}^{+} & =\prod_{r'_{2}=r_{2},r'_{1}}^{\prime} e^{i\pi\left(\hat{n}_{\vec{r}',1}+\hat{n}_{\vec{r}',3}+\hat{n}_{\vec{r}'+e_{1},2}+\hat{n}_{\vec{r}'+e_{1},4}\right)},\\
\hat{F}_{r_{2}}^{-} & =\prod_{r'_{2}=r_{2},r'_{1}}^{\prime} e^{i\pi\left(\hat{n}_{\vec{r}',2}+\hat{n}_{\vec{r}',4}+\hat{n}_{\vec{r}'+e_{2},1}+\hat{n}_{\vec{r}'+e_{2},3}\right)},
\end{align}
\end{subequations}
where $\prod^{\prime}$ means that $r_1^{\prime}$ runs over the values of the same odevity as $r_2$ in the product. Note that $\hat{F}_{r_{2}}^{+}=\hat{F}_{r_{2}+1}^{-}$. 
The $Z_2$ flux operator on the edge octagon plaquette becomes
\begin{equation}
\phi_{o}=\left\{
\begin{array}{ll}
\hat{F}_{r_{2}}^{+}\hat{F}_{r_{2}}^{-}\hat{D}_{\left(L_{1},r_{2}\right),3}\hat{D}_{\left(2,r_{2}\right),1}, & r_{2}=odd, \\
\hat{F}_{r_{2}}^{+}\hat{F}_{r_{2}}^{-}\hat{D}_{\left(L_{1}-1,r_{2}\right),3}\hat{D}_{\left(1,r_{2}\right),1}, & r_{2}=even.
\end{array}
\right.
\end{equation}
It is easy to verify that all $\hat{F}_{r_{2}}^{+}$ are commute with each other and with the Hamiltonian and $(\hat{F}_{r_{2}}^{+})^{2}=1$. 
However, $\hat{D}_{\vec{r},\mu}$ anticommutes with $\hat{F}_{r_{2}}^{+}$ and $D_{\vec{r},\mu}$ is no longer a good quantum number under the periodic boundary condition. 
Insteadly, we choose $\{\phi_{t},\phi_{o},\Phi_1,\Phi_2 \}$ as a set of good quantum numbers, 
where $\Phi_1=F_{r_{2}=1}^{+}$ and $\Phi_2=\prod_{r_{1}=1,r_{2}=odd}D_{\vec{r},1}$ are the global $Z_2$ fluxes along the $\hat{1}$- and $\hat{2}$-direction respectively. 
More discussions on this can be found in\cite{Mott}. Numerically we find that the ground states are all $\pi$-flux states as well under the periodic boundary condition. 
The ground states are of $Z_2\times Z_2$ topologically degenerate characterized by the global fluxes $\Phi_1=\pm 1$ and $\Phi_2=\pm 1$.  

For the four topologically degenerate ground states on a torus, the Hamiltonian is translational invariant. 
Define the four-component Majorana spinor $\Gamma_{\vec{r}}=(\gamma_{\vec{r},1}, \gamma_{\vec{r},2},  \gamma_{\vec{r},3},  \gamma_{\vec{r},4})^{T}$,
we can perform the Fourier transformation $\Gamma_{\vec{r}}=\frac{1}{\sqrt{N}}\sum_{\vec{q}}e^{i\vec{q}\cdot\vec{r}}\Gamma_{\vec{q}}$ to diagonalize the Hamiltonian,
where $N=L_{1}\times L_{2}$ is the number of the unit cells and the wave vector $\vec{q}$ lies in the first Brillouin zone.
Note that all the components in $\Gamma_{\vec{r}}$, say, $\gamma_{\vec{r},\mu}$, are Majorana fermions, 
the relation $\Gamma_{\vec{q}}^{\dagger}=\Gamma_{-\vec{q}}$ should be satisfied in the Fourier transformation, although $\Gamma_{\vec{q}}$ is no longer a Majorana spinor.
The Hamiltonian in the reciprocal space reads $H=\frac{i}{2}\sum_{\vec{q}}\Gamma_{\vec{q}}^{\dagger}h\left(\vec{q}\right)\Gamma_{\vec{q}}$, with
\begin{equation}
h\left(q\right)=\left(\begin{array}{cccc}
0 & J^{z}e^{iq_2} & -J^{xy}_1e^{\frac{i}{2}q_1} & J^{xy}_2e^{-\frac{i}{2}q_1}\\
 & 0 & J^{xy}_2e^{-\frac{i}{2}q_1} & -J^{xy}_1e^{\frac{i}{2}q_1}\\
 &  & 0 & -J^{z}e^{iq_2}\\
-h.c. &  &  & 0
\end{array}\right),
\end{equation}
where $\hat{1}=\left(1,0\right)$ and $\hat{2}=\left(0,1\right)$ are two unit vectors.
To diagonalize the Hamiltonian, we employ the Bogoliubov transformation and the diagonalized form reads $H=\sum_{\vec{q}s}E_{\vec{q}s}\zeta_{\vec{q}s}^{\dagger}\zeta_{\vec{q}s}$, 
where $\zeta_{\vec{q}s}$ is the Bogoliubov quasiparticle and $E_{\vec{q}s}$ is the quasiparticle energy dispersion with the band indices $s=1,2,3,4$. From the energy dispersions
\begin{equation}
E_{\vec{q}s}=\pm\frac{1}{2}\sqrt{J^{2}\pm2J_{2}^{xy}\sqrt{\left(\cos\left(q_{x}\right)J_{1}^{xy}\right)^{2}+\left(\cos\left(q_{y}\right)J^{z}\right)^{2}}},
\end{equation}
where $J^{2}=\left(J_{1}^{xy}\right)^{2}+\left(J_{2}^{xy}\right)^{2}+\left(J^{z}\right)^{2}$, 
we find the system is gapful except $\left(J_{2}^{xy}\right)^{2}=\left(J_{1}^{xy}\right)^{2}+\left(J^{z}\right)^{2}$, when a nodal point in the spectrum appears at $\vec{q}=(0,0)$.
The energy dispersion near the nodal point is linear.  
Thus we conclude that the system has two gapful phases separated by the critical line $\left(J_{2}^{xy}\right)^{2}=\left(J_{1}^{xy}\right)^{2}+\left(J^{z}\right)^{2}$, which is a circle on the $2D$ space spanned by the two ratios of coupling constants $J_{1}^{xy}/J_{2}^{xy}$ and $J^{z}/J_{2}^{xy}$.

\begin{figure}[hptb]
\begin{center}
\includegraphics[width=8.4cm]{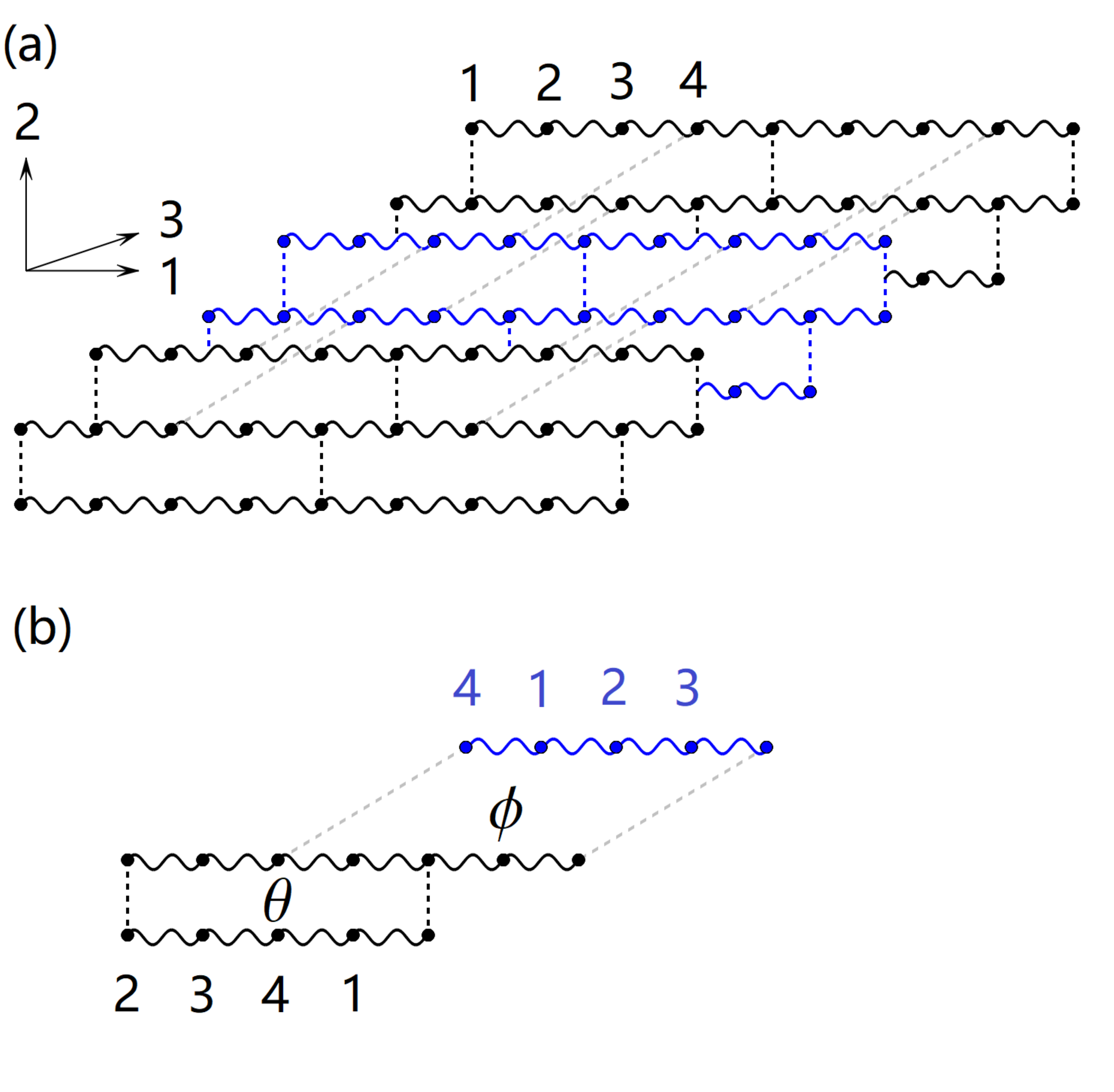}
\end{center}
\caption{(color online) (a) A lattice hosting the $3D$ $xy$ bond model, which is topologically equivalent to hyperhoneycomb lattice. 
Wave lines denote local $xy$-bonds, and dashed lines denote local $z$-bonds. 
Each unit cell contains four sites labeled as $\mu=1,2,3,4$. To denote different layers we use black and blue dots, nevertheless these dots should be understood as the white sites in previous sections.
(b) $\theta$ and $\phi$ are the $Z_2$ flux operators defined on two types of elementary decagon plaquettes.
}
\label{fig:3dxy2}
\end{figure}

{\em Example II: $3D$ $xy$ bond model.}
We choose the $3D$ $xy$ bond model (see FIG.~\ref{fig:3d}(a)) as a representative model in $3D$ and study it through the exact solution.
This model shares the same topologically equivalent hyperhoneycomb lattice with the one studied in Refs.\cite{mandal,Kimchi,Trebst15,Hermanns,Nasu}, but possesses different spin interactions.
The original hyperhoneycomb has three types of local bonds, $x$-bonds, $y$-bonds and $z$-bonds, but all the $x$-bonds and $y$-bonds are replaced by $xy$-bonds in our $3D$ $xy$ bond model. 
The model Hamiltonian is given by
\begin{eqnarray}\label{eq:3dhspin}
H & = &\sum_{\vec{r}}J^{xy}\sigma^{x}_{\vec{r},1}\sigma^{y}_{\vec{r},2}+J^{xy}\sigma^{x}_{\vec{r},3}\sigma^{y}_{\vec{r},4}\nonumber \\
  &   &+J^{xy}\sigma^{x}_{\vec{r},2}\sigma^{y}_{\vec{r},3}+J^{xy}\sigma^{x}_{\vec{r},4}\sigma^{y}_{\vec{r}+e_{1},1}\nonumber \\
  &   & +J^{z}\sigma^{z}_{\vec{r},1}\sigma^{z}_{\vec{r}+e_{2},2} +J^{z}\sigma^{z}_{\vec{r},3}\sigma^{z}_{\vec{r}+e_{3},4},
\end{eqnarray}
where the unit cell consists of four sites and is denoted as $\left(\vec{r},\mu\right)$ with $\vec{r}=(r_1,r_2,r_3)$ the Bravais vector and $\mu=1,2,3,4$ the sublattice indices,
as illustrated in FIG.~\ref{fig:3dxy2}(a).
The three Bravais basis are chosen as $e_{1}=\left(1,0,0\right)$, $e_{2}=\left(1,0,0\right)$ and $e_{3}=\left(0,0,1\right)$. 
$J^{xy}$ is the $xy$-bond coupling constant along the $\hat{1}$-direction, and $J^{z}$ is the $z$-bond coupling constant along $\hat{2}$- and $\hat{3}$-direction.
By the Jordan-Wigner transformation, Eq.(\ref{eq:3dhspin}) is fermionized as 
\begin{eqnarray}\label{eq:3dhmajorana}
H & = &\sum_{\vec{r}}-iJ^{xy}\gamma_{\vec{r},1}\gamma_{\vec{r},2}-iJ^{xy}\gamma_{\vec{r},3}\gamma_{\vec{r},4}\nonumber \\
  &   &-iJ^{xy}\gamma_{\vec{r},2}\gamma_{\vec{r},3}-iJ^{xy}\gamma_{\vec{r},4}\gamma_{\vec{r}+e_{1},1}\nonumber \\
  &   &+iJ^{z}D_{\vec{r},1}\gamma_{\vec{r},1}\gamma_{\vec{r}+e_{2},2} +iJ^{z}D_{\vec{r},3}\gamma_{\vec{r},3}\gamma_{\vec{r}+e_{3},4},
\end{eqnarray}
where $D_{\vec{r},1}=-i\eta_{\vec{r},1}\eta_{\vec{r}+e_2,2}$ and $D_{\vec{r},3}=-i\eta_{\vec{r},3}\eta_{\vec{r}+e_3,4}$.
It is obvious that $D_{\vec{r},1}^2=D_{\vec{r},3}^2=1$.

As mentioned, under the open boundary condition, all the $\hat{D}_{\vec{r},\mu}$'s commute with each other and with the Hamiltonian, 
and form a set of good quantum numbers. So that they can be replaced by their eigenvalues $\pm{}1$. 
There are two types of elementary plaquettes, say, decagon plaquettes, on the hyperhoneycomb lattice as shown in FIG.~\ref{fig:3dxy2}(b)
and we define the $Z_{2}$ flux operators on each types of decagon plaquettes as follows, 
\begin{subequations}
\begin{equation}
\begin{split}
\hat{\theta}_{\vec{r}} &= -\sigma^{y}_{\vec{r},1}\sigma^{z}_{\vec{r},2}\sigma^{z}_{\vec{r},3}\sigma^{z}_{\vec{r},4}\sigma^{x}_{\vec{r}+e_{1},1}\\
&\quad\times{}\sigma^{x}_{\vec{r}+e_{12},2}\sigma^{z}_{\vec{r}+e_{12},1}\sigma^{z}_{\vec{r}+e_{2},4}\sigma^{z}_{\vec{r}+e_{2},3}\sigma^{y}_{\vec{r}+e_{2},2}\\
&=\hat{D}_{\vec{r},1}\hat{D}_{\vec{r}+e_{1},1},
\end{split}
\end{equation}
and
\begin{equation}
\begin{split}
\hat{\phi}_{\vec{r}} &= -\sigma^{y}_{\vec{r},3}\sigma^{z}_{\vec{r},4}\sigma^{z}_{\vec{r}+e_{1},1}\sigma^{z}_{\vec{r}+e_{1},2}\sigma^{x}_{\vec{r}+e_{1},3}\\
&\quad\times{}\sigma^{x}_{\vec{r}+e_{13},4}\sigma^{z}_{\vec{r}+e_{13},3}\sigma^{z}_{\vec{r}+e_{13},2}\sigma^{z}_{\vec{r}+e_{13},1}\sigma^{y}_{\vec{r}+e_{3},4}\\
&=\hat{D}_{\vec{r},3}\hat{D}_{\vec{r}+e_{1},3},
\end{split}
\end{equation}
\end{subequations}
where $e_{12}=e_{1}+e_{2}$ and $e_{13}=e_{1}+e_{3}$. It is easy to see that $\hat{\theta}_{\vec{r}}^2=\hat{\phi}_{\vec{r}}^2=1$. 
$\hat{\theta}_{\vec{r}}$ and  $\hat{\phi}_{\vec{r}}$  commute with each other and with the Hamiltonian and can be replaced by their eigenvalues $\theta_{\vec{r}}=\pm 1$ and $\phi_{\vec{r}}=\pm 1$. 
Then, we numerically find that the ground states are all zero-flux states, i.e., $\theta_{\vec{r}}$ = $\phi_{\vec{r}}$ = 1 everywhere, which is the same as the Kitaev honeycomb model.

For periodic boundary condition, similar to the $2D$ model, extra boundary terms will be introduced due to Jordan-Wigner transformation. 
We define the following operators 
\begin{equation}
\hat{F}_{r_{2},r_{3}}=\prod_{r'_{2}=r_{2},r'_{3}=r_{3},r'_{1}}e^{i\pi\sum_{\mu}\hat{n}_{\vec{r}',\mu}},
\end{equation}
which counts the parity of the fermion numbers for each chain along the $\hat{1}$-direction. Then the boundary terms involved in the Hamiltonian read
\begin{equation}
\begin{split}
&J^{xy}\sigma^{x}_{(L_{1},r_{2},r_{3}),4}\sigma^{y}_{(1,r_{2},r_{3}),1}\\
=& -iJ^{xy}\hat{F}_{r_{2},r_{3}}\gamma_{(L_{1},r_{2},r_{3}),4}\gamma_{(1,r_{2},r_{3}),1},
\end{split}
\end{equation}
where $L_{1}$ is the lattice length in $\hat{1}$-direction.
Meanwhile the $Z_{2}$ flux operators on the edge decagon plaquette become 
\begin{equation}
\begin{split}
&\hat{\theta}_{(L_{1},r_{2},r_{3})}=\hat{D}_{(L_{1},r_{2},r_{3}),1}\hat{D}_{(1,r_{2},r_{3}),1}\hat{F}_{r_{2},r_{3}}\hat{F}_{r_{2}+1,r_{3}},\\
&\hat{\phi}_{(L_{1},r_{2},r_{3})}=\hat{D}_{(L_{1},r_{2},r_{3}),3}\hat{D}_{(1,r_{2},r_{3}),3}\hat{F}_{r_{2},r_{3}}\hat{F}_{r_{2},r_{3}+1}.
\end{split}
\end{equation}
For the same reason as in the $2D$ example, $\hat{D}_{\vec{r},\mu}$ anticommutes with $\hat{F}_{r_{2},r_{3}}$ such that ${D}_{\vec{r},\mu}$ can not be chosen as good quantum numbers under the periodic boundary condition. 
However, we can still choose $Z_{2}$ fluxes $\{\theta_{\vec{r}},\phi_{\vec{r}},\Phi_{1}, \Phi_{2},\Phi_{3}\}$ as the set of good quantum numbers, 
where $\Phi_1$, $\Phi_2$ and $\Phi_3$ are the eigenvalues of the global $Z_2$ flux operators $\hat{\Phi}_{1}=\hat{F}_{1,1}$, $\hat{\Phi}_{2}=\prod_{r_{1}=r_{3}=1,r_{2}}\hat{D}_{\vec{r},1}$ and
$\hat{\Phi}_{3}=\prod_{r_{1}=r_{2}=1,r_{3}}\hat{D}_{\vec{r},1}$, which are defined along the $\hat{1}$-direction, $\hat{2}$-direction and $\hat{3}$-direction respectively. 
These operators all commute with each other and with the Hamiltonian and are idempotent. 
Numerically we find the ground states are zero flux as well as those under open boundary condition. 
In the thermodynamic limit, the ground states are of $Z_2 \times Z_2 \times Z_2$ topological degeneracy characterized by $\Phi_{1}=\pm 1$, $\Phi_{2}=\pm 1$ and $\Phi_{3}=\pm 1$.

Now we study the bulk excitations on top of one of the zero flux ground states through the Fourier transformation. 
Note that the energy dispersions will shift by $(\pm\pi/L_1,\pm\pi/L_2,\pm\pi/L_3)$ among these degenerate ground states on a $L_1\times L_2 \times L_3$ torus. 
The Hamiltonian in the reciprocal space reads $H=\frac{i}{2}\sum_{\vec{q}}\Gamma_{\vec{q}}^{\dagger}h\left(\vec{q}\right)\Gamma_{\vec{q}}$ with 
\begin{equation}\label{eq:3dhamk}
h\left(q\right)=\left(\begin{array}{cccc}
0 & -J^{xy}+J^{z}e^{iq_2} & 0 & J^{xy}e^{-iq_1}\\
 & 0 & -J^{xy} & 0\\
 &  & 0 & -J^{xy}+J^{z}e^{iq_3}\\
-h.c. &  &  & 0
\end{array}\right).
\end{equation}
Without loss of generality, we choose both $J_{xy}$ and $J_{z}$ to be non-negative. By diagonalizing Eq.\eqref{eq:3dhamk}, 
we find that the system has (1) a gapped phase when $J_{z}>2J_{xy}$ and (2) a gapless phase when $J_{z}<2J_{xy}$, which are separated by the critical point $J_{z}=2J_{xy}$.
In a gapless state with $J_{z}<2J_{xy}$, a nodal ring appears with linear energy dispersion along the directions perpendicular to the ring.  
At the critical point $J_{z}/J_{xy}=2$, the nodal ring shrinks to a gapless point at $\vec{q}=(0,0,0)$ and energy dispersion around this nodal point is linear along all the directions. 
This nodal ring is protected by the time reversal symmetry and a perturbation term, such as the Zeeman splitting $\sum_{\vec{r},\mu}\vec{h}\cdot\vec{\sigma}_{\vec{r},\mu}$ with $\vec{h}\propto(1,1,1)$,
will open a gap along the nodal ring except at two singular points, which are so-called Weyl points.
Similar results were obtained in the hyperhoneycomb model in Ref.\cite{mandal,Trebst15,Hermanns}.

\section{Summary and Discussions}

In summary, we construct a class of exactly solvable generalized Kitaev spin-$1/2$ models in arbitrary dimensions.
The basic idea is to construct exactly solvable spin chains and couple them to form a connected diagram. 
The allowed spin interactions include two-spin interactions as well as multiple-spin interactions.
The construction is subjected to two elementary rules and several supplementary rules.
The Jordan-Wigner transformation is employed to prove the exact solvability, by which a constructed spin model can be mapped to a gas of Majorana fermions coupled to static $Z_2$ gauge fields. 
We classify the exactly solvable models according to their parent models. All the other models can be generated by some operations from the parent models.
It is noted that there exist a dual model to each exactly solvable spin model in this class.     
As two examples, a $2D$ square-octagon model and a $3D$ $xy$ bond model are demonstrated. 

Finally, we would like to make some comments on the construction and possible realization: 
(1) With $xy$-bonds and $yx$-bonds introduced, this class of exactly solvable models are beyond the category of compass model\cite{nussinov2013}. 
(2) The translational symmetry or periodicity is not a necessary condition to the exact solvability. Thus one is able to map an exactly solvable spin model with randomness to a free fermion model with random hopping and pairing.  
(3) We specify the $z$-bonds as the ``glue" bonds to couple spin chains under the construction. It is for the convenience to adopt the Jordan-Wigner transformation. By global spin rotations, other corresponding spin models can be constructed as well.
(4) These constructed models could be realized in cold atom systems and coordination polymers. Moreover, it is possible to tailor our model Hamiltonians for specific materials.

\section{Acknowledgement.}

JJM is supported by Postdoctoral Science Foundation of China (No.119103S284).
YZ is supported by National Key Research and Development Program of China (No.2016YFA0300202),
National Basic Research Program of China (No.2014CB921201), National Natural Science Foundation of China (No.11774306), the Key Research Program of the Chinese Academy of Sciences (Grant No. XDPB08-4) and the Fundamental Research Funds for the Central Universities in China.
FCZ is supported by NSFC grant 11674278, National Basic Research Program of China (No. 2014CB921203), and the CAS Center for Excellence in Topological Quantum Computation.

\end{document}